# Ligand Equilibrium Influences Photoluminescence Blinking in $CsPbBr_3$: A Change Point Analysis of Widefield Imaging Data

Shaun Gallagher[1,†], Jessica Kline[1,†], Farzaneh Jahanbakhshi[2], James C. Sadighian[1], Ian Lyons[1], Gillian Shen[1], Andrew M. Rappe[2], and David S. Ginger[1,*]

[1]Department of Chemistry, University of Washington, Seattle, WA 98195, USA

[2]Department of Chemistry, University of Pennsylvania, Philadelphia, Pennsylvania 19104, USA

* Corresponding author: dginger@uw.edu

[†]These authors contributed equally to this work

**Abstract:**

Photoluminescence intermittency remains one of the biggest challenges to realizing perovskite quantum dots (QDs) as scalable single photon emitters. We compare $CsPbBr_3$ QDs capped with different ligands, lecithin, and a combination of oleic acid and oleylamine, to elucidate the role of surface chemistry on photoluminescence intermittency. We employ widefield photoluminescence microscopy, sampling the blinking behavior of hundreds of QDs. Using change point analysis, we achieve the robust classification of blinking trajectories, and we analyze representative distributions from large numbers of QDs ($N_{lecithin}$ = 1308, $N_{oleic\ acid/oleylamine}$ =1317). We find that lecithin suppresses blinking in $CsPbBr_3$ QDs compared to oleic acid/oleylamine. Under common experimental conditions, lecithin-capped QDs are 7.5 times more likely to be non-blinking and spend 2.5 times longer in their most emissive state, despite both QDs having nearly identical solution photoluminescence quantum yields. We measure photoluminescence as a function of dilution and show that the differences between lecithin and oleic acid/oleylamine capping emerge at low concentrations during preparation for single particle experiments. From experiment and first principles calculations, we attribute the differences in lecithin and oleic acid/oleylamine performance to differences in their ligand binding equilibria. Consistent with our experimental data, density functional theory calculations suggest a stronger binding affinity of lecithin to the QD surface compared to oleic acid/oleylamine, implying a reduced likelihood of ligand desorption during dilution. These results suggest that using more tightly binding ligands is a necessity for surface passivation and consequently, blinking reduction in perovskite QDs used for single particle and quantum light experiments.

**Introduction**

Inorganic cesium lead bromide ($CsPbBr_3$) perovskite quantum dots (QDs) are promising solution-processable materials for a wide range of optoelectronic applications.[1] These materials exhibit high (>90%) photoluminescence quantum yields (PLQY),[2] narrow ensemble photoluminescence linewidths,[3] and emission spectra that can be tuned throughout the visible region.[4] These properties have motivated increased efforts to use perovskite QDs as the active layer in devices such as light-emitting diodes (QLEDs),[5] photovoltaics,[6] and even X-ray detectors.[7] More recently, their high degree of quantum coherence has positioned $CsPbBr_3$ QDs as leading candidates for next-generation quantum light sources - scalable, coherent single photon emitters[8–11]

Successful single photon emitters must demonstrate a high degree of single photon purity, have long coherence times, and be deterministically positioned within nanophotonic cavities.[12] Undesirable characteristics for single emitter candidates include photobleaching, particle heterogeneity, spectral diffusion, and photoluminescence intermittency.[12] Prior reports for $CsPbBr_3$ QDs have shown high single photon purity,[13,14] quantum interference between sequential photons,[15] and progress towards deterministic cavity positioning.[16–18] However, as even the highest quality perovskite QDs currently exhibit photoluminescence intermittency and spectral diffusion, events which have been linked in other quantum dot systems;[12,19] these challenges must be overcome to further improve these systems and reduce their photoluminescence linewidth.[8,20,21]

Quantum-confined materials possess high surface-area-to-volume ratios, meaning their properties are heavily influenced by their surfaces as has been extensively explored for II-VI quantum dots.[22–27] Improving $CsPbBr_3$ single photon emitters requires a more fundamental understanding and optimization of their surface chemistry.[28] $CsPbBr_3$ QDs are often described as defect tolerant and can achieve narrow linewidths and high PLQYs without the use of a core-shell heterostructure (as in II-VI or III-VI QDs). However as defect formation is still a thermodynamically favorable process, $CsPbBr_3$ QDs require an organic ligand shell for both colloidal stability and surface defect passivation.[29,30] Due to the highly ionic bonding character of $CsPbBr_3$ QDs, many popular ligand systems contain both positive and negative charges, either in the form of zwitterions or monodentate ligand pairs.[31,32] The identity of this ligand layer plays a major role in modulating the properties of $CsPbBr_3$ QDs, including PLQY,[33] linewidth,[34] device performance,[35–39] and colloidal stability.[31,32]

Sustained synthetic efforts have identified several promising ligand species for $CsPbBr_3$ QDs. These include monodentate ligand pairs, like oleic acid/oleylamine[40], quaternary amines, like didodecyldimethylammonium bromide[41,42] and multi-dentate ligands, such as sulfobetaine,[43] lecithin,[44] diquaternary amines,[34] and phosphonic acids.[45,46] Despite ligands resulting in measurably different ensemble properties, including long-term colloidal stability,[44] and surface-defect induced electronic traps[37] there has been a lack of emphasis on understanding how these ligands affect the properties of single QDs. While some multi-dentate ligands display improved ensemble linewidths and near unity PLQY,[43,44] results from device integration are mixed,[38] underscoring the need for application-specific investigations, particularly at the single-particle level.

One way to probe the effects of surface chemistry on the optical properties of individual QDs is through monitoring their photoluminescence intermittency, or blinking.[47,48] Like other single photon emitters,[12] $CsPbBr_3$ QDs exhibit time-dependent variations in photoluminescence intensity at the single emitter level.[49] This variation in photoluminescence intensity arises from fluctuations in the non-radiative decay rates.[49] Prior studies of blinking in perovskite QDs have focused on understanding and passivating defects – through temperature-dependent measurements[50], and in-situ halide treatments.[51] However, relatively few studies have examined how varying the surface ligand chemistry of CsPbBr$_3$ QDs affects their blinking behavior.

Here we study the blinking dynamics of CsPbBr$_3$ QDs prepared via two different synthetic routes that result in dots passivated with two different ligands, a typical oleic acid/oleylamine based hot injection resulting in oleic acid/oleylamine-capped dots, and a recently reported room-temperature slow growth synthesis, leading to lecithin-capped dots via ligand exchange.[52] We utilize widefield photoluminescence microscopy and implement an automated particle selection and unique change point analysis algorithm adapted to the noise profile of scientific cameras. This approach allows for significantly improved throughput and more robust sampling of the QD distributions. We find that, while both syntheses produce QDs with comparable PLQYs as synthesized, the lecithin-capped dots show dramatically reduced blinking when prepared and studied at the single particle level. We show that this difference primarily arises from the difference in ligand capping, and, despite similar ensemble characteristics, lecithin passivation leads to a dramatic decrease in blinking, seen through an improved ON percent distribution, a larger ON/OFF ratio, and a larger non-blinking fraction. We reconcile the apparent discrepancy between the ensemble photoluminescence properties and those of single dots by showing that the oleic acid/oleylamine dots suffer from significant degradation of their photoluminescence properties upon dilution to appropriate concentrations for single particle experiments. We present experimental evidence and theoretical analysis using density functional theory (DFT) calculations that rationalize these differences in terms of the ligand binding affinity, with lecithin showing significantly stronger binding to the QD surface compared to oleic acid/oleylamine.

**Results and Discussion**

We synthesized CsPbBr$_3$ QDs using one of two different methods from the literature to produce QDs with either oleic acid/oleylamine[40] or lecithin[52] as surface ligands (Figure 1a). Figures 1c and 1d show the UV-Vis absorption and photoluminescence spectra for representative synthetic batches of oleic acid/oleylamine-capped, and lecithin-capped dots, respectively. We find that the two ligand compositions impart similar ensemble properties to the QDs. The band-edge absorbance onset is nearly identical for both sets of QDs. However, lecithin-capped QDs show distinctly sharper excitonic features due to their spheroidal shape. The lecithin dots exhibit an emission at 506 nm with a FWHM of 19 nm, while the photoluminescence of oleic acid/oleylamine-capped QDs is slightly broader, as expected from the hot-injection synthesis, with an emission maximum at 509 nm and a 24 nm FWHM (Figures 1c and 1d). The lecithin-capped QDs batches tend to have slightly higher PLQY (95 ± 2% compared to 90 ± 5%) but shorter lifetimes (4.43 ± 0.04 ns vs 4.95 ± 0.02 ns) than oleic acid/oleylamine-capped QDs (Figure 1b-d). The shorter lifetime is qualitatively consistent with predictions for a spherical crystal habit, which was predicted to have faster recombination due to difference in dielectric confinement and asymmetry of the photon's electric field between spherical and cubic QDs at comparable sizes.[53] Interestingly, the spheroidal lecithin-capped QDs also show a low-energy tail in the photoluminescence lineshape, consistent with prior reports.[52,54] From the absorption band edge,[55] the oleic acid/oleylamine-capped QDs are determined to be 9.5 ± 2.7 nm, comparable in size to our 9.7 ± 1.5 nm lecithin-capped QDs, which were measured via TEM (Figure S1).

Studies of QD blinking statistics are commonly limited by two factors: the quantity of observed QDs and the analysis method.[56,57] Using confocal microscopy, blinking traces must be collected in series; limiting typical sample distributions to tens of QDs.[58] Instead, we collect blinking traces in parallel using widefield photoluminescence microscopy, allowing rapid collection of thousands of blinking traces.[59] Perovskite QDs also display complex blinking statistics, including multi-level dynamics, beyond simple two-level ON/OFF dynamics.[60] A data set of this size and complexity requires an automated, robust analysis method to extract accurate statistics. We satisfy this requirement through change point analysis (CPA) adapted to the noise profile of scientific cameras.[61–63] This combination of widefield imaging and CPA provides more representative sampling than previously possible, enabling a more accurate description of intermittency in these systems.

Figure 2 provides an overview of our analysis method on a representative image sequence of lecithin-capped $CsPbBr_3$ QDs. After the collection of a widefield image sequence, our analysis method works to extract the time series describing each QD's behavior and completes an unbiased classification of the photoluminescence trajectory. Figure 2a shows the mean intensity image of a widefield image sequence. Single QDs (shown circled in red and labeled with a white number) are identified by an automated particle picking algorithm, designed to select small bright spots from a dark background.[64] Figure 2b shows the time series behavior of representative QDs (2, 6, and 11) extracted from the same widefield image sequence as Figure 2a. We generate these time series by intensity averaging the central pixels of selected QDs at each image in the sequence. Figures 2c and 2d show the photoluminescence trajectories identified for QD 2 (2-level) and QD 6 (3-level) respectively. We use change point analysis[61,62], which applies Bayesian statistics to find the photoluminescence trajectory which best describes a time series and to assign states to the photoluminescence vs. time trajectories. The methods section contains additional details about the analysis. Figure S2 shows representative trajectories from the oleic acid/oleylamine-capped quantum dots.

To classify the blinking dynamics of lecithin-capped QD samples systematically, we analyzed the dynamics of five separate synthetic batches of comparable quality (Table S1 and Figure S3) to control for batch-to-batch variation, which can often serve as a confounding variable.[65] Figures S4 and S5 explore the statistics for these five synthetic batches in more detail. The blinking statistics of lecithin-capped QDs we report in the main text of this paper are aggregated from the blinking behavior of all five batches.

We focus on four key metrics to compare blinking statistics: the non-blinking fraction, the ON percentage, the number of CPA levels in each trajectory, and the weighted ON/OFF ratio. We define the non-blinking fraction as all the QDs which are in their highest intensity level for greater than 95% of the observation time, consistent with previously reported widefield blinking studies.[48] The ON percentage quantifies how long we observed each QD in its highest intensity level state during the observation window. The weighted ON/OFF ratio quantifies, given an infinite observation time frame, how much more likely a QD is to be in its most emissive intensity level. Further details explaining how we calculate these statistics from our CPA results are in the supporting information. Here, we compare the blinking statistics across five different synthetic batches of lecithin-capped QDs (1308 QDs in total), to an oleic acid/oleylamine synthetic batch (1317 QDs in total). Additional comparisons between other synthetic batches of lecithin-capped and oleic acid/oleylamine-capped QDs are shown in Figure S6.

Surprisingly, despite similar ensemble properties, at the single particle level, we observe disparate blinking behaviors from the two QD compositions. Figure 3b shows a similar distribution of identified CPA levels between the two compositions. Nevertheless, at the single particle level lecithin-capped QDs significantly outperform those capped by oleic acid/oleylamine. Lecithin-capped QDs exhibit a larger non-blinking fraction (0.15 vs 0.02, Figure 3a), higher average ON percentage (68% vs 30%, Figure 3c), and a larger weighted ON/OFF ratio (3.1 vs 0.73, Figure 3d); all metrics are consistent with dramatic reductions in their blinking. Additional metrics comparing blinking in lecithin-capped and oleic acid/oleylamine-capped $CsPbBr_3$ QDs are shown in Figure S7. At face value, this result seems surprising – after all, the ensemble is the sum of the individual single particles, yet the single particle data differ dramatically.

To better understand these contrasting blinking distributions, we turn to concentration dependent photoluminescence and time-resolved photoluminescence (TRPL) studies. First, we use TRPL for a comparative estimate of surface quality based on lifetime duration. At the QD concentration used to acquire our ensemble PLQY data the oleic acid/oleylamine-capped QDs have an average lifetime of 4.95 ± 0.02 ns and a stretching exponent of $\beta = 0.779 \pm 0.001$ (Figure 4a). After diluting the QD sample to the concentration used to prepare single particle samples, we find that the average lifetime has decreased to 2.08 ± 0.03 ns with a stretching exponent of $\beta = 0.455 \pm 0.002$ (Figure 4a). The observed decrease in average lifetime and stretching exponent indicates that the surface passivation and sample homogeneity have worsened during the dilution process.[66] In contrast, Figure 4b shows the average lifetime for lecithin-capped QDs remains essentially constant during the dilution process (4.43 ± 0.04 ns vs 4.50 ± 0.08 ns). Additional concentration dependent lifetimes for oleic acid/oleylamine-capped and lecithin-capped QDs are shown in Figures S8 and S9.

We next turn to concentration-dependent photoluminescence studies for a more quantitative understanding of the changes in surface passivation inferred from the TRPL data. If the PLQY of a material remains constant with dilution, the integrated photoluminescence intensity should scale linearly with concentration. A sub-linear trend would indicate a loss of PLQY with dilution. This fact allows integrated photoluminescence intensity to serve as a proxy for PLQY at concentrations too dilute to be directly measured in the integrating sphere. Figure 4c plots the photoluminescence intensity versus concentration for both the lecithin-capped and oleic acid/oleylamine-capped QDs. For the lecithin-capped dots, we observe a linear relationship between the photoluminescence intensity and concentration over three orders of magnitude in concentration, spanning from typical measurement concentrations for solution PLQY to a concentration range below that used in our single QD microscopy experiments. In contrast, Figure 4c shows that the integrated photoluminescence intensity for the oleic acid/oleylamine-capped QDs is linear at higher concentrations, but falls off rapidly as the concentration decreases. This additional decrease in intensity is consistent with the change in the non-radiative recombination dynamics we deduce from the TRPL shown in Figure 4a. Approximately midway between the PLQY concentration (indicated with a green square) and the single QD microscopy concentration (indicated with a green circle), oleic acid/oleylamine-capped QDs see a significant decrease in integrated photoluminescence intensity, which we interpret as a decrease in PLQY due to ligand desorption.

Previous studies on a range of QDs have attributed photoluminescence decreases during washing to an equilibrium between the ligand bound to the QD surface and free-ligand in solution which shifts to favor free-ligand during dilution or antisolvent washing.[41] As the concentration of unbound ligand decreases during dilution, the equilibrium shifts, and monodentate ligands like oleic acid and oleylamine are likely to detach from perovskite QDs in two ways – as a neutral ligand pair (OAm-OA) or as ligand-ion pairs (OA$^-$-Cs$^+$ and OAm$^+$-Br$^-$) which leave behind additional vacancies.[31] As the ligands desorb from the surface, the number of surface traps increases which can explain both the sub-linear photoluminescence trend and increased blinking observed in oleic acid/oleylamine-capped QDs. These single particle results are consistent with prior reports of decreased PLQY in halide perovskite QDs at low concentrations.[67]

To test whether ligand desorption might cause the observed non-linear decrease in quantum dot emission with decreasing concentration, we investigate the effect of the adding back additional ligand to dilute solutions of oleic acid/oleylamine-capped QDs. Figure 4d plots the photoluminescence intensity of a dilute solution oleic acid/oleylamine-capped QDs as a function of excess ligand addition. The solution's photoluminescence intensity increases after adding small amounts of oleylamine or lecithin, resulting in a maximum 120% increase in photoluminescence intensity. The photoluminescence recovery associated with the addition of other ligands is shown in Figure S10.

Taken together the results above suggest that the increased blinking behavior in the oleic acid/oleylamine-capped QDs stems from ligand loss with dilution. We speculate that the lack of full photoluminescence recovery (which is expected to be approximately 220% according to the trend line) after excess ligand addition might indicate that ligand loss in oleic acid/oleylamine-capped $CsPbBr_3$ contributes to the irreversible degradation of a fraction of these QDs. These results are consistent with widefield photoluminescence images of lecithin-capped and oleic acid/oleylamine-capped samples prepared at the same concentration indicating that a smaller fraction of oleic acid/oleylamine capped QDs remain emissive after dilution (Figure S11).

To rationalize the differences in blinking behavior between oleic acid/oleylamine- and lecithin-capped QDs, we study the quantum dot electronic structure and compute the binding energies of the capping ligands to the surface of the QDs. To begin, we calculated the density of states for the pristine and the defective (containing $v•_{Br}$ or $v'_{Cs}$) slabs of $CsPbBr_3$, shown in Figure S12. We performed DFT-PBE calculations considering spin orbit coupling effects, which are known to strongly influence the position of band edges in lead halide perovskites.[68] We calculated the band gaps of the pristine slab as well as $v•_{Br}$ and $v'_{Cs}$ containing slabs of $CsPbBr_3$ to be 1.4 eV, 1.17 eV and 1.0 eV, respectively. Figures S12b and S12c depict near gap states due to the presence of surface vacancies ($v•_{Br}$ and $v'_{Cs}$) that are known potential recombination centers,[69] and the band gap closure that stems from them compared to the pristine slab (Figure S12a). We posit that the origin of the difference in blinking behavior between lecithin- and oleic acid/oleylamine- capped QDs lies in the ability of the ligands to passivate these defects. To this end, we investigated the potential difference in surface binding energy between lecithin and oleic acid/oleylamine using first principles studies.

We conducted a comparative study on the passivation of $v'_{Cs}$-$v•_{Br}$ vacancy pairs by oleic acid/oleylamine and lecithin ligands. Figure 5a shows the orthorhombic phase of $CsPbBr_3$ and different types of the adjacent (types *i, ii* and *iii*) and non-adjacent (type *iv*) vacancy pairs used in our calculations. Figures 5b and 5c depict the monodentate ligand pair and the bidentate ligand binding the QD surface and passivating an adjacent surface vacancy pair. As we do not expect the long alkyl tails of the ligands to participate in surface binding, we use simplified forms of the ligands (methylamine, propionic acid and truncated lecithin, Figure S13 for computational purposes. Given the dynamic equilibrium between the charged and neutral ligands in the solution, we studied surface defect passivation for both cases. Between charged and neutral ligands, we found that $OAm^+$/$OA^-$ passivating an adjacent vacancy pair is the most stable configuration and that the neutral ligand pair could only passivate a non-adjacent vacancy pair and exhibits a lower binding affinity (-1.07 eV) than that of the charged ligands (-1.95 eV). Additionally, our calculations revealed that the neutral ligand pair binding the defective surface in the presence of solvent is relatively thermodynamically unfavorable (Table S2). Figure 5d shows the binding energies calculated at the DFT level for lecithin and $OAm^+$/$OA^-$ (-3.22 eV and -1.95 eV, respectively), when passivating the vacancy pairs. Figure 5d and Table S3 summarize the binding energies calculated for the different vacancy pair types and show that for both adjacent and non-adjacent vacancies, lecithin binds the QD surface more strongly than $OAm^+$/$OA^-$. However, for non-adjacent vacancies, the relative binding energy of $OAm^+$/$OA^-$ increases while the lecithin binding energy decreases. For both ligands we rationalize this difference by considering the larger distance between the non-adjacent vacancies (3.7-5.0 Å vs 8.5 Å). For $OAm^+$/$OA^-$, the larger distance between the two vacancies results in a weaker intermolecular interaction and a subsequent higher affinity for the QD surface. Lecithin, as a bidentate ligand, demonstrates a preference for passivating adjacent vacancy pairs. Our theoretical findings suggest that lecithin binds more strongly to the QD surface, which aligns qualitatively with experiments. However, the magnitudes of calculated binding energies do not reproduce the observed ligand equilibrium shifting to unbound ligands over the experimental concentrations. Such observed difference led us to consider the impact of solvents that appear critical in representing the experimental conditions.

Since experiments indicate that the dilution effects on ligand binding equilibria should be significant, we further refined the binding energies by including solvation effects (Figure S14). As shown in Figure 5e, the net change in the binding energy due to solvation effects is remarkable for both ligands in the representative case of a type *i* vacancy pair. The calculated binding energy of -0.26 eV for $OAm^+/OA^-$ suggests a high likelihood of ligand desorption which we speculate to be the root cause of the observed difference in the blinking behavior between oleic acid/oleylamine- and lecithin-capped $CsPbBr_3$ QDs. However, the binding energy of lecithin remains significantly large (more negative) upon solvation and can explain the observed reduction in blinking. Notably, the binding energy for $OAm^+/OA^-$ calculated *via* DFT (-0.26 eV) is in good agreement with the analogous binding energy estimated from Figure 4c (-0.34 eV). We ascribe the remaining discrepancies between the calculated and experimental binding energies to the absence of entropy and finite-temperature effects in our calculations, the inherent challenges in accurately modelling solvation effects, and the limitations in the assumptions behind the experimental analysis (see supporting information). The supporting information contains a detailed explanation of solvation effects calculations, equilibrium constants and binding energies summarized in Table S4.

**Conclusion**

We use widefield photoluminescence microscopy and CPA to analyze the blinking statistics of 2,600 QDs to compare the effect of bidentate and monodentate ligands on photoluminescence intermittency in $CsPbBr_3$. We show that, despite similar ensemble properties, using lecithin as a capping ligand results in QDs with a non-blinking fraction 7.5 times larger than that of the oleic acid/oleylamine ligand pair. This difference in performance is explained through a ligand binding equilibrium where oleic acid and oleylamine desorb during serial dilutions, degrading the surface. This ligand binding model is supported by two key sets of findings. First, concentration dependent photoluminescence measurements, which expose a non-linear relationship between oleic acid/oleylamine photoluminescence intensity and concentration. Second, theoretical investigation and binding energy calculations at the DFT level highlight striking differences in solvated-binding energy between lecithin and $OAm^+/OA^-$. This blinking suppression highlights the promise of these lecithin capped $CsPbBr_3$ QDs as scalable single photon emitters.

We also demonstrate the promising capabilities of widefield photoluminescence microscopy paired with CPA. Together, these two techniques facilitate rapid, representative sampling of QD blinking statistics, allowing more accurate determinations of QD behavior and its connections to surface chemistry. This method can compare blinking through the number of intensity levels, the non-blinking fraction, the percent of time spent in the most intense state, the expected dwell time in an ON state, the expected dwell time in an OFF state and the weighted ON/OFF ratio. The variety of statistics available from this analysis method, combined with the larger sample size available *via* widefield microscopy allow us to determine representative blinking statistics of a sample more accurately, enabling deeper and more accurate investigations of the role QD surface chemistry plays in blinking.

While we address the difference in blinking caused by using oleic acid/oleylamine and lecithin as ligands, it is also possible that the spheroidal shape of these QDs influences their blinking as well. Though we have demonstrated, through several control experiments, the significant impact ligand binding has on photoluminescence blinking, it is still possible that the spheroidal shape of these QDs contributes to their better stability. Given the interest in these materials for such a wide range of applications and their impressive non-blinking characteristics, a more detailed understanding of how the spherical crystal habit affects trap density and the ligand-binding equilibrium could prove valuable. Future work is needed to explore this relationship. The differences in blinking behavior observed here also lead naturally to the exploration of other ligands used to passivate $CsPbBr_3$, including alternative zwitterions,[70] dications,[34] and tightly binding monodentate ligands.[29]

High throughput of widefield imaging, combined with the robust nature of CPA, permits detailed analysis of larger sample sizes of QDs. As we have demonstrated, the choice of surface ligand is vital to ensure that the QDs maintain their properties at the low concentrations required for both single particle characterization and nanophotonic cavity integration. The experimental techniques discussed here pave the way for systematic investigations into the root causes of photoluminescence intermittency in QDs.

**Methods**

*Chemicals*

Cesium carbonate ($Cs_2CO_3$, 99.9% metals basis), octadecene (ODE, 90%), oleic acid (OA, ≥ 99%), oleylamine (OAm, 70%), diisooctylphosphinic acid (DOPA, 90%), trioctylphosphine oxide (TOPO, 99%), hexanes (anhydrous 95%), hexanes (≥ 95%), polystyrene (PS, MW 280,000), toluene (anhydrous 99.8%), octane (anhydrous ≥ 99%), acetone (90%), isopropyl alcohol (IPA, 90%) and acetonitrile (ACN, anhydrous 99.8%) were purchased from Millipore Sigma. Lecithin (90%) and lead (II) bromide ($PbBr_2$, 99.998% metals basis) were purchased from Alfa Aesar. ODE and OAm were degassed by freeze-pump-thaw method and stored in a glovebox at 0°C. $Cs_2CO_3$ was dried in a vacuum oven at 120°C for 8 hours and stored in a glovebox. All other chemicals were used as received.

*Synthesis of oleic acid/oleylamine-capped $CsPbBr_3$ QDs*

Oleic acid/oleylamine-capped $CsPbBr_3$ were synthesized according to Protesescu et al[40] and washed with ACN according to Zhang et al.[33] After washing, the resulting QDs were dissolved in 2 mL anhydrous hexanes, filtered through a 0.45 μm PTFE filter and stored at 0°C in a glovebox.

*Synthesis of Lecithin-Capped $CsPbBr_3$ QDs*

The precursor solutions for lecithin-capped $CsPbBr_3$ were prepared according to Akkerman et al.[52] The QDs were synthesized under ambient conditions in 6 mL hexanes with 160 μL of 0.04 M $PbBr_2$-TOPO, 320 μL of 0.2 M TOPO, 80 μL of 0.02 M Cs-DOPA. After 30 minutes of stirring at room temperature 80 μL of 0.13 M lecithin was added to the reaction. The QDs were subsequently washed by adding acetone in a 3:1 v:v ratio and centrifuged at 10,000 rpm for 5 minutes. The supernatant was discarded, and the precipitate was dissolved in 2 mL anhydrous hexanes, filtered through a 0.45 μm PTFE filter and stored at 0°C in a glovebox.

*Ensemble Characterization*

Absorbance spectra of the QD solutions were performed using a Perkin-Elmer Lambda 950 UV/Vis/NIR Spectrometer with a range of 400-700 nm with an integration time of 0.5 s.

Steady-state photoluminescence spectra were acquired using Perkin-Elmer Fluorescence Spectrometer LS 55. Spectra were collected using an excitation wavelength of 405 nm. The emission bandwidth was kept at 3 nm and dwell time at 0.2 s.

Time resolved photoluminescence measurements at 405 nm excitation were acquired using a commercial PicoQuant FluorTime 100 system with LDH-405 laser diode, a 405 nm picosecond pulsed diode laser. The repetition rate is controlled by an external trigger input from a PicoHarp PDL 800-B laser driver and was set to 5 MHz. A photomultiplier tube (PMT) detector was used in TCSPC mode with an instrumental response function (IRF) of approximately 400 ps. The instrument response function (IRF) was measured via laser scatter from a solution of colloidal silica (LUDOX). Concentration dependent photoluminescence was collected on the same instrument with a laser repetition rate of 50 MHz. TRPL lifetime analysis was performed using a custom Python IRF-reconvolution package. Lifetimes were fit using a stretch exponential decay (Equation 1).[66]

$$I(t) = A \exp\left(-\frac{t}{\tau_k}\right)^\beta + C \qquad (1)$$

Where *A* is the pre-exponential factor, $\tau_k$ is the lifetime of the decay, *C* is the background of the measurement and *β* is the distribution of decay rates. $\tau_{str}$, the average lifetime of a stretch exponential, is calculated according to Equation 2 where Γ is the gamma function.[66]

$$\tau_{str} = \frac{\tau_k}{\beta}\Gamma\left(\frac{1}{\beta}\right) \qquad (2)$$

The series of QD solutions prepared for TRPL and concentration dependent TRPL are described in Table S5.

Photoluminescence quantum yield measurements (PLQY) were performed on a commercial integrating sphere system (Hamamatsu Photonics K.K). PLQY values are determined using a white light source (Hamamatsu Mercury Xenon Lamp) and a monochromator for wavelength selection (405 nm) as the excitation source to illuminate the samples in an integrating sphere (Hamamatsu Photonics K.K). The optical density of samples was kept below 0.1 at the excitation wavelength to minimize reabsorption effects. Spectral correction was performed using a calibrated white light source (Ocean Insight HL-3P-INT-CAL) to correct for the responsivity of the detector. PLQY was calculated using the following formula:

$$PLQY = \frac{I_{em,\,sample} - I_{em,\,blank}}{I_{ex,\,blank} - I_{ex,\,sample}} * 100 \qquad (3)$$

Where $I_{em,sample}$ and $I_{em,blank}$ are the integrated area under the curve in the emission region (450-600 nm) of the sample and the neat hexane blank, respectively. The $I_{ex,sample}$ and $I_{ex,blank}$ are the integrated area under the curve in the excitation region (395-415 nm) of the sample and the neat hexane blank respectively.

*Widefield Microscopy Sample Preparation*

Low fluorescence glass coverslips (VistaVision #1.5 22x22 mm, VWR) were cleaned through a sequential sonication of 10 minutes in soap solution, de-ionized water (x2), acetone and IPA. Each side of the coverslips were ozone cleaned for 23 minutes before spin coating.

Low concentration QD solutions were prepared in 5 wt% PS in toluene via serial dilutions. For oleic acid/oleylamine-capped QDs 2 μL of stock QD solution were added to 2 mL of PS-toluene (solution A$_{oleic\ acid/oleylamine}$). 200 μL of solution A$_{oleic\ acid/oleylamine}$ were added to 2 mL of PS-toluene (solution B$_{oleic\ acid/oleylamine}$). For lecithin-capped QDs 1 μL of stock QD solution were added to 2 mL of PS-toluene (solution A$_{lec}$). 100 μL of solution A$_{lec}$ were added to 2 mL of PS-toluene (solution B$_{lec}$). 100 μL of solution B$_{lec}$ were added to 2 mL of PS-toluene (solution C$_{lec}$). Single QDs films were prepared via spin coating solution B$_{oleic\ acid/oleylamine}$ (60 μL) or solution C$_{lec}$ (60 μL) onto the cleaned coverslips at 2,000 rpm for 40 seconds. The films were stored in a glovebox until measurement.

*Widefield Microscopy*

Widefield microscopy measurements were performed on a Nikon TE2000 inverted optical microscope using a CFI Super Fluor 40x Oil immersion objective (NA = 1.3), with Olympus F immersion oil. The illumination source was a 415 nm LED (SOLIS-415C, Thor Labs) at a power density of 9 mW/cm$^2$. The following filters were used for the measurement: ET510/80m (Chroma), FF01-424/SP-25 (Semrock) and ZT442rdc (Chroma) mounted in Chroma Laser TIRF for Nikon TE2000/T filter cube. Videos of the sample photoluminescence in time were collected on a Prime 95B (Photometrics) camera for 6 minutes and 40 seconds (8000 total images) with an integration time of 50 ms per image. Individual QD blinking traces were extracted from the videos using a custom-built Python package for selecting bright objects from a dark background.[64] Briefly, individual quantum dots were selected using the Laplacian of Gaussian method to automatically identify bright single particles from the mean image. The image coordinates of these identified particles were used to extract the relevant single particle time traces from the image sequence for further analysis. The efficacy of particle identification was assessed by constricting the microscope field of view to rule out false positive selections (Figure S15).

*Analysis of Blinking Traces Using Change Point Analysis*

We used a version of Change Point Analysis (CPA) adapted for a Gaussian-distributed time series. Our home-built Python CPA package is based on the equations published by Yang et al.[62] A more detailed explanation of our CPA package can be found in the supporting information – including a confusion matrix quantifying the performance of our CPA package on synthetically generated blinking data (Figure S16) and a detailed breakdown of how we extract our main blinking statistics from CPA fit traces. Our particle selection and CPA code is publicly available at: https://github.com/GingerLabUW/Widefield-CPA.

*Computational Methods*

To investigate the passivation of different types of adjacent and non-adjacent vacancies, nonstoichiometric Cs-terminated slabs with a vacuum of ~15 Å were constructed from 2x1, 1x2, and 2x2 replications (corresponding to vacancy-pair types i-ii, iii and iv, respectively) of the PBE-relaxed conventional lattice of orthorhombic $CsPbBr_3$ (8.19 Å, 8.54 Å, 11.99 Å), shown in Figure 5a. DFT calculations were performed using the Quantum Espresso suite of codes,[71] whereby energetically minimized atomic structures were obtained using the Perdew–Burke–Ernzerhof (PBE) formulation.[72] Ultrasoft pseudopotentials were used to describe the interaction between the valence electrons and the ionic cores. Kohn–Sham orbitals were expanded in a plane wave basis set with a kinetic-energy cutoff of 60 Ry and a density cutoff of 420 Ry. Van der Waals interactions were considered by applying the empirical D3 dispersion correction scheme of Grimme.[73] Band gaps and partial density of states were calculated with PBE[72], including spin-orbit coupling (SOC) effect,[68] that is known to strongly influence the position of band edges in lead-containing halide perovskites. Although PBE+SOC is not expected to reproduce the experimental gap of this material, we applied SOC to ensure that the band edges and the trap states near them were captured properly. An explicit solvent model was employed to simulate solvation effects, providing a more realistic representation that, unlike implicit models, includes direct and specific interactions between the solvent and solute. A detailed explanation of these calculations can be found in the supporting information.

**Author Contributions**

The manuscript was written through contributions of all authors. All authors have given approval to the final version of the manuscript.

**Funding**

This work, and the roles of S.G., J.K. J.C.S., I.L., G.S., F.J., A.M.R., and D.S.G. were primarily supported by the National Science Foundation under the STC IMOD Grant (No. DMR-2019444). The authors acknowledge the use of facilities and instruments at the Photonics Research Center (PRC) at the Department of Chemistry, University of Washington, as well as that at the Research Training Testbed (RTT), part of the Washington Clean Energy Testbeds system. Part of this work was carried out at the Molecular Analysis Facility, a National Nanotechnology Coordinated Infrastructure site at the University of Washington which is supported in part by the National Science Foundation (NNCI-1542101), the Molecular Engineering & Sciences Institute, and the Clean Energy Institute. Computational support was provided by the National Energy Research Scientific Computing Center (NERSC), a US Department of Energy, Office of Science User Facility, located at Lawrence Berkeley National Laboratory, operated under contract no. DE-AC02-05CH11231.

The authors declare no competing financial interests.

### Acknowledgements

S.G. and J.K. would like to acknowledge Margherita Taddei (graduate student, University of Washington) for scientific discussions related to photoluminescence lifetime measurements, Muammer Yaman (graduate student, University of Washington) for his particle identification code and Seth Dale (graduate student, Colorado School of Mines) for his insights into improving the runtime of our change point analysis code. We also thank Hannah Even for her contribution during her stay in the Rappe group as part of the Research Experience for Undergraduates program supported by IMOD.

**Figures:**

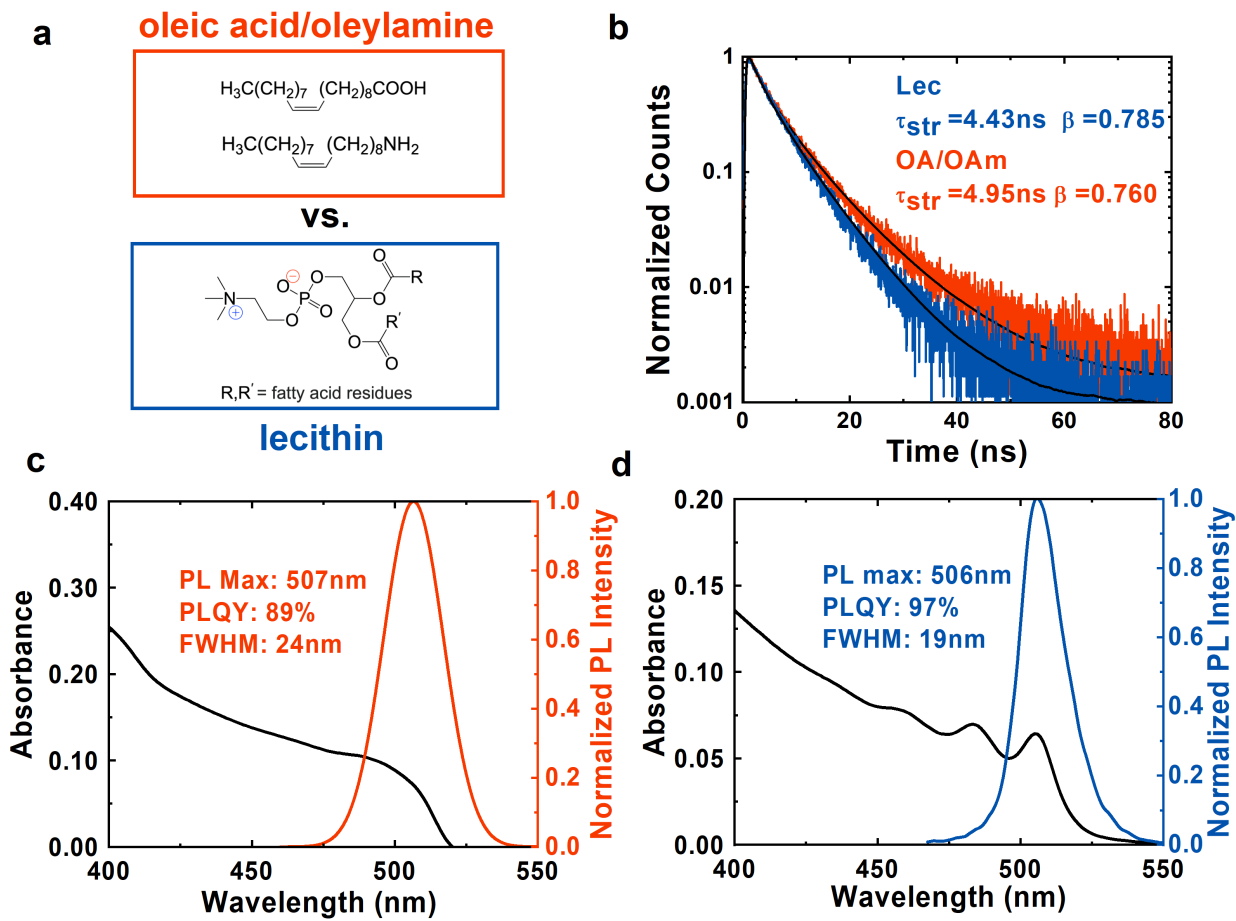

**Figure 1. Ensemble characterization of lecithin and oleylamine/oleic acid capped QDs: a)** Chemical structures of zwitterionic lecithin (blue) and the monodentate ligand pair oleic acid (OA) and oleylamine (OAm) (red). **b)** Ensemble lifetimes of oleic acid/oleylamine-capped and lecithin-capped $CsPbBr_3$ QDs fit to a stretch exponential function (Equation 1) **c)** Ensemble solution characterization of oleic acid/oleylamine-capped $CsPbBr_3$ QDs **d)** Ensemble solution characterization lecithin-capped $CsPbBr_3$ QDs.

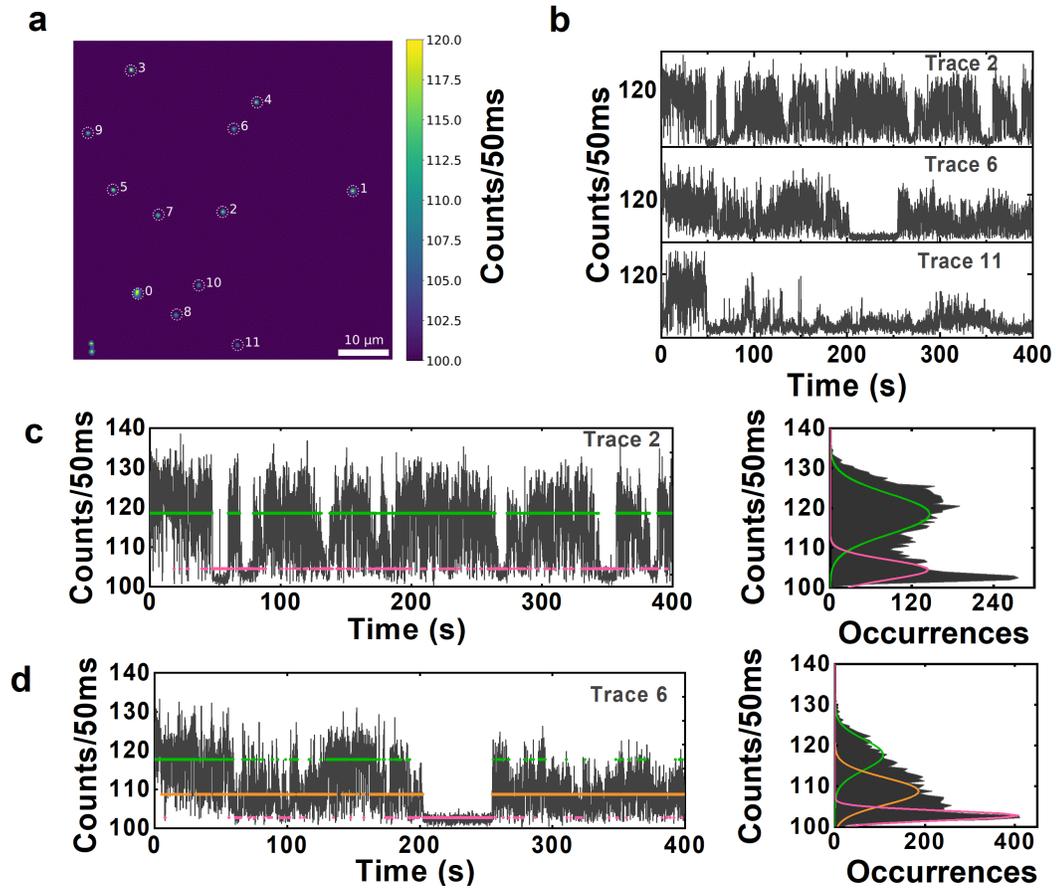

**Figure 2. Workflow of widefield photoluminescence microscopy measurements. a)** Automated particle selection results; particle selections are plotted on the mean intensity image in the sequence. For clarity, we show results from a small region of the overall image. Red circles labeled with a white number indicate quantum dot locations. **b)** Photoluminescence time series extracted from the selected regions in 2a corresponding to identified QDs 2, 6 and 11. **c)** CPA fitting of the time series from QD 2 (a). CPA finds that this time series is best described by a photoluminescence trajectory containing two average intensity levels (pink and green). **d)** CPA fitting of the time series from QD 6 (a). CPA finds that this time series is best described by a photoluminescence trajectory containing three average intensity levels (pink, orange and green).

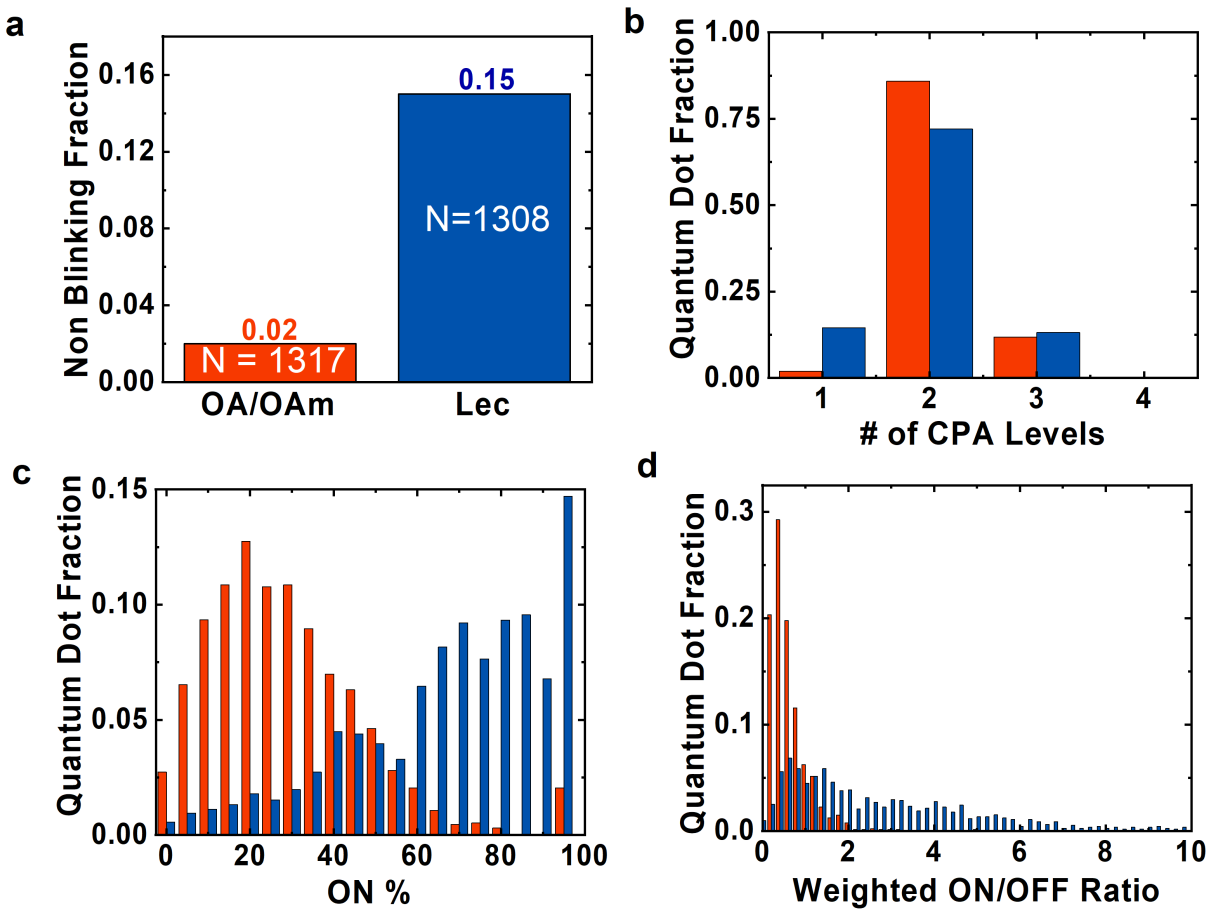

**Figure 3. Comparison of blinking statistics between oleic acid/oleylamine-capped (OA/OAm, red) and lecithin-capped (Lec, blue) CsPbBr$_3$. a)** The non-blinking fraction (>95% ON). **b)** The distribution of the number of intensity levels fit by CPA across all studied QDs. **c)** Distribution of the ON percentages for all QDs. **d)** The weighted ON/OFF ratio distribution.

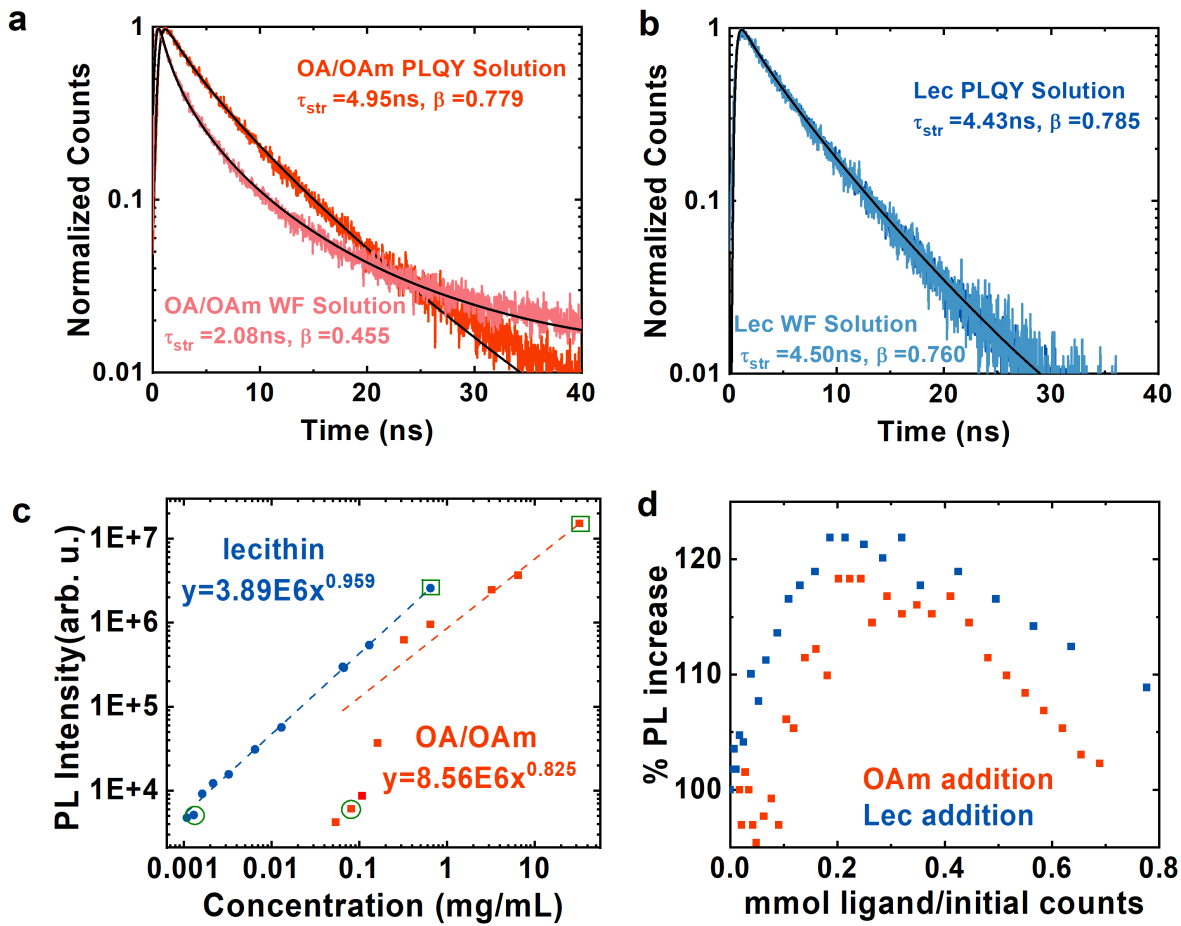

**Figure 4. Rationalizing reduced blinking in lecithin-capped QDs**. **a)** Concentration dependent TRPL for oleic acid/oleylamine-CsPbBr$_3$ at the PLQY concentration (green box in (c)) and widefield (WF) concentration (green circle in (c)). **b)** Concentration dependent TRPL for lecithin-capped CsPbBr$_3$ at the PLQY concentration (green box in (c)) and widefield concentration (green circle in (c)). **c)** Concentration dependent photoluminescence intensity for oleic acid/oleylamine (OA/OAm, red) and lecithin (blue) capped QDs. Intensity vs. concentration values are fit to a power law (a power law exponent of 1 indicates a linear relationship). The experimental concentrations for PLQY and widefield photoluminescence blinking measurements on both compositions are indicated with green box and a green circle, respectively. **d)** Photoluminescence increase resulting from the addition of oleylamine (OAm) and lecithin (Lec) to a widefield concentration solution (green circle in (c)) of oleic acid/oleylamine-capped CsPbBr$_3$ QDs.

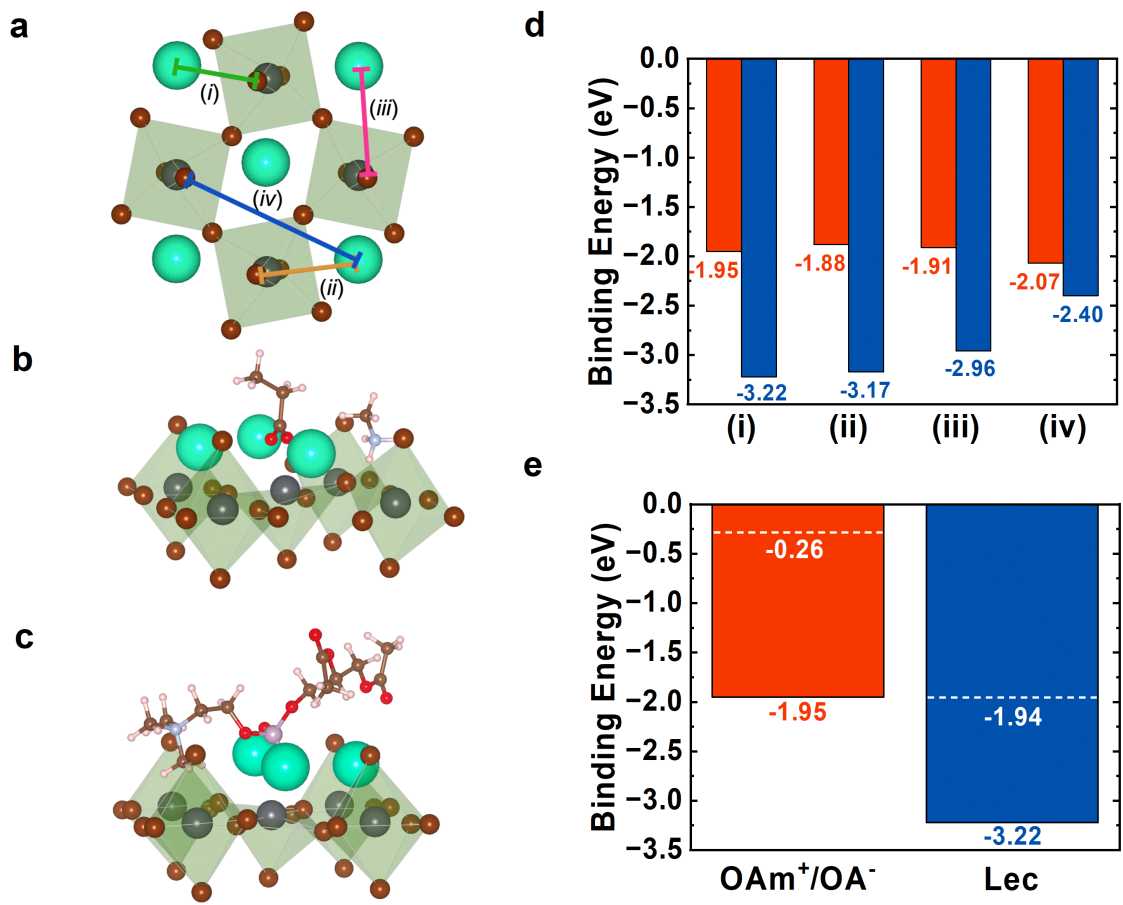

**Figure 5.** Calculated binding energies for binary monodentate ligand system oleic acid/oleylamine (OAm$^+$/OA$^-$) and bidentate lecithin (Lec) **a)** Top view of the PBE-relaxed lattice of orthorhombic CsPbBr$_3$ (8.19 Å, 8.54 Å, 11.99 Å) and different types of vacancy pairs studied here marked as *i, ii, iii* and *iv* **b)** PBE+D3-relaxed CsPbBr$_3$ slab with a type *i* vacancy pair passivated by oleic acid/oleylamine **c)** PBE+D3-relaxed CsPbBr$_3$ slab with a type *i* vacancy pair passivated by lecithin **d)** Calculated binding energies in eV for CsPbBr$_3$ slabs containing v'$_{Cs}$ and v˙$_{Br}$ passivated by oleic acid/oleylamine (red) and lecithin (blue), at PBE+D3 level of theory. **e)** Vacancy pair type *i* binding energy for both lecithin and oleic acid/oleylamine with binding energy in the presence of solvent represented by white dashed lines.

# Supporting Information

## Ligand Equilibrium Influences Photoluminescence Blinking in CsPbBr$_3$: A Change Point Analysis of Widefield Imaging Data

Shaun Gallagher[1†], Jessica Kline[1†], Farzaneh Jahanbakhshi[2], James C. Sadighian[1], Ian Lyons[1], Gillian Shen[1], Andrew M. Rappe[2], and David S. Ginger[1*]

[1]Department of Chemistry, University of Washington, Seattle, WA 98195, USA

[2]Department of Chemistry, University of Pennsylvania, Philadelphia, Pennsylvania 19104, USA

* Corresponding author: dginger@uw.edu

[†]These authors contributed equally to this work

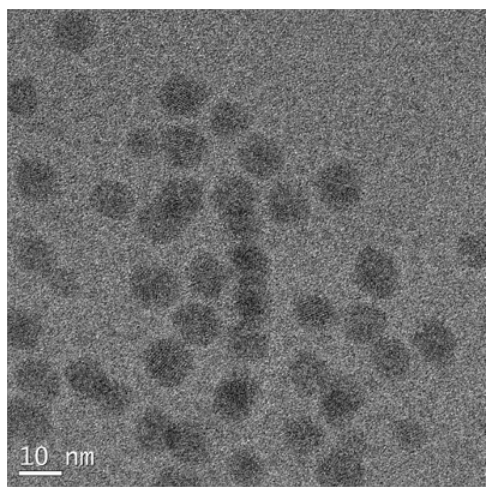

Figure S1. TEM of lecithin quantum dots showing spherical crystal habit

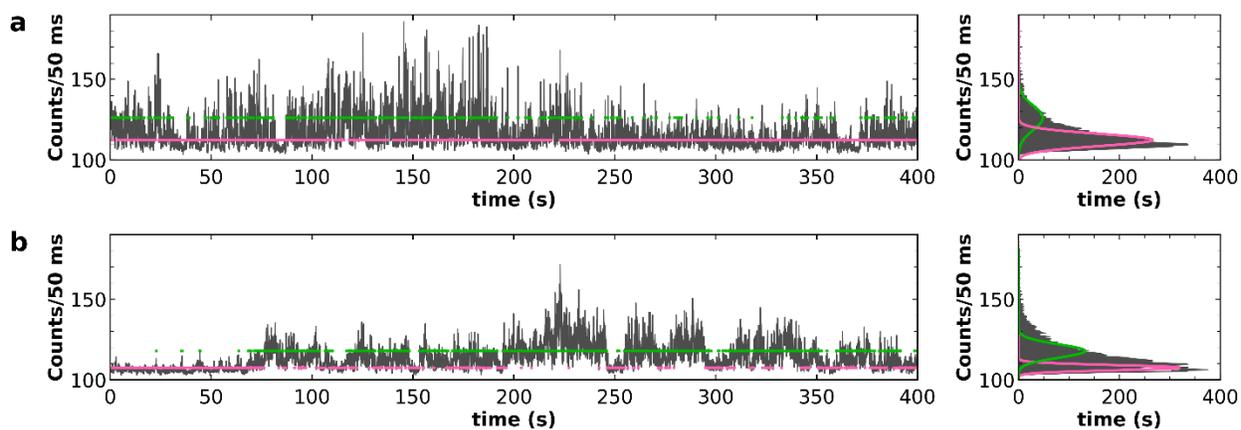

**Figure S2.** Randomly selected example blinking traces from OA/OAm-QDs showing the CPA identified intensity trajectories. Both traces have been fit to two average intensity-levels (pink and green).

Table S1. Ensemble characteristics of the five lecithin synthetic batches used in this study

|  | **Absorption Band Edge (nm)** | **PL Maximum (nm)** | **PL Linewidth (nm)** | **PLQY (%)** |
|---|---|---|---|---|
| **LEC 1** | 503 | 505 | 19 | 97 |
| **LEC 2** | 505 | 508 | 19 | 97 |
| **LEC 3** | 505 | 507 | 17 | 98 |
| **LEC 4** | 507 | 509 | 16 | 94 |
| **LEC 5** | 507 | 508 | 18 | 93 |

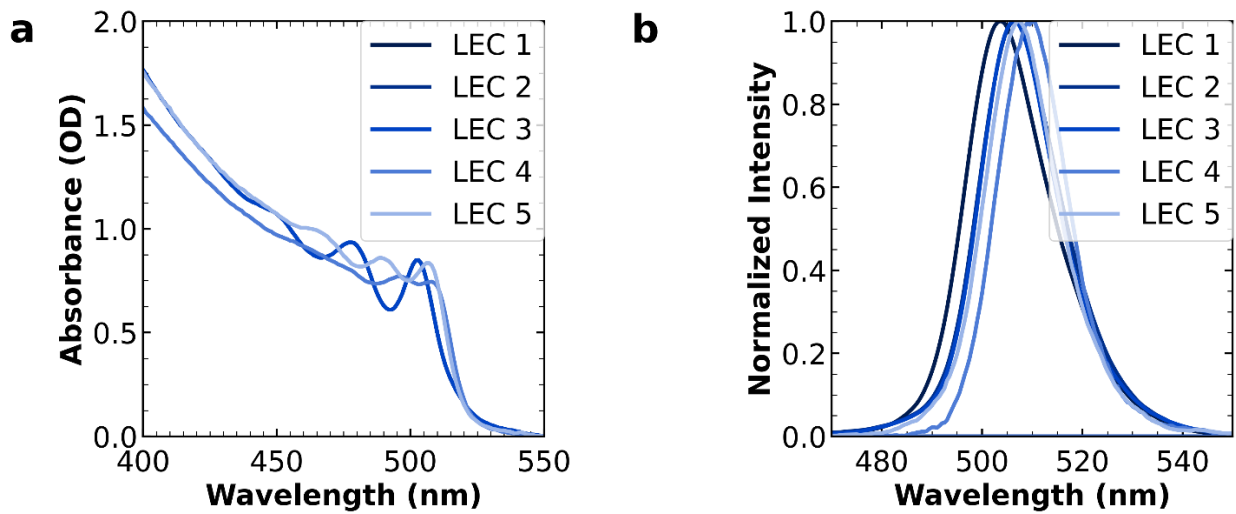

Figure S3. **a)** absorption spectra of the five lecithin synthetic batches used in this study **b)** PL of five lecithin synthetic batches used in this study

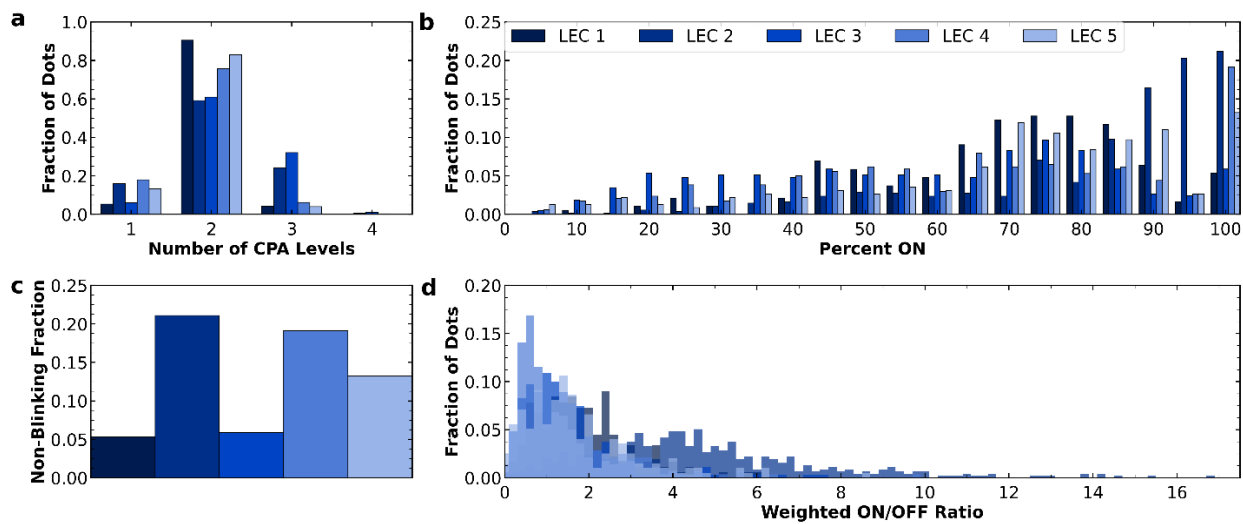

Figure S4. Comparisons of blinking behavior between the five lecithin synthetic batches across our four major blinking metrics. These statistics are comprised of 188, 553, 374, 340 and 227 QDs for batches 1-5 respectively **a)** number of intensity levels fit to each blinking trace by CPA. The average CPA levels are 2.0 ± 0.3, 2.1 ± 0.6, 2.2 ± 0.6, 1.9 ± 0.5 and 1.9 ± 0.4 for batches 1-5 respectively **b)** percent of the trace duration that each QD spends ON. The average ON percentages are 66 ± 17%, 77 ± 20%, 53 ± 24%, 56 ± 23% and 64 ± 22% for batches 1-5 respectively **c)** non-blinking fraction of QDs in each batch, 0.053, 0.22, 0.059, 0.19 and 0.13 for batches 1-5 respectively and **d)** the weighted ratio of expected time spent ON to expected time spent OFF

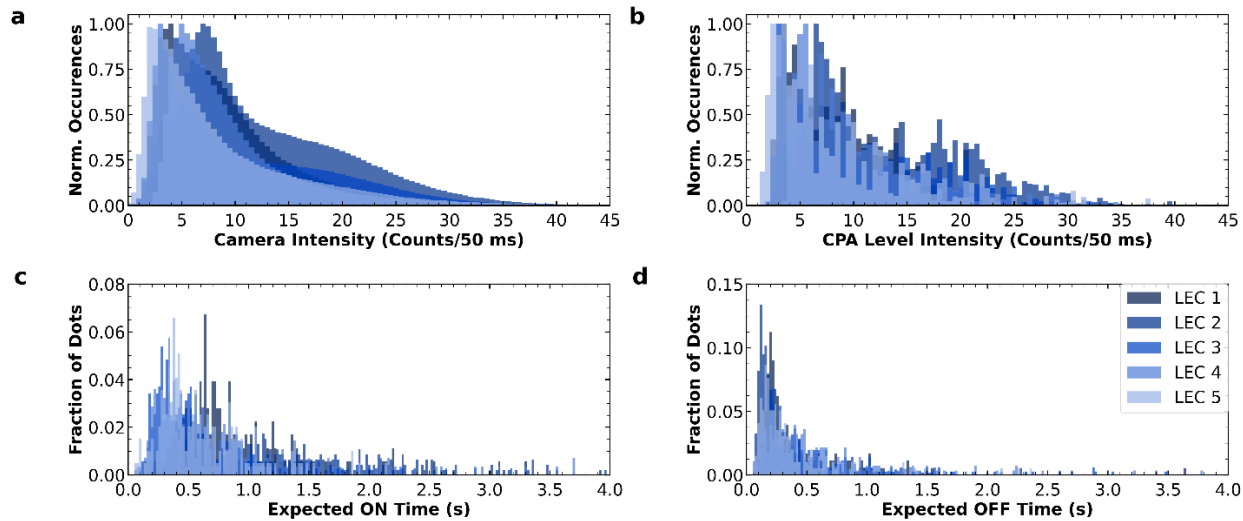

Figure S5. Comparison of blinking behavior between the five lecithin synthetic batches across four additional metrics **a)** distribution of intensities in the blinking traces according to the camera **b)** distribution of intensities in the blinking traces according to the CPA fits **c)** distribution of the expected ON times from power-law dynamics and **d)** distribution of the expected OFF times from power-law dynamics.

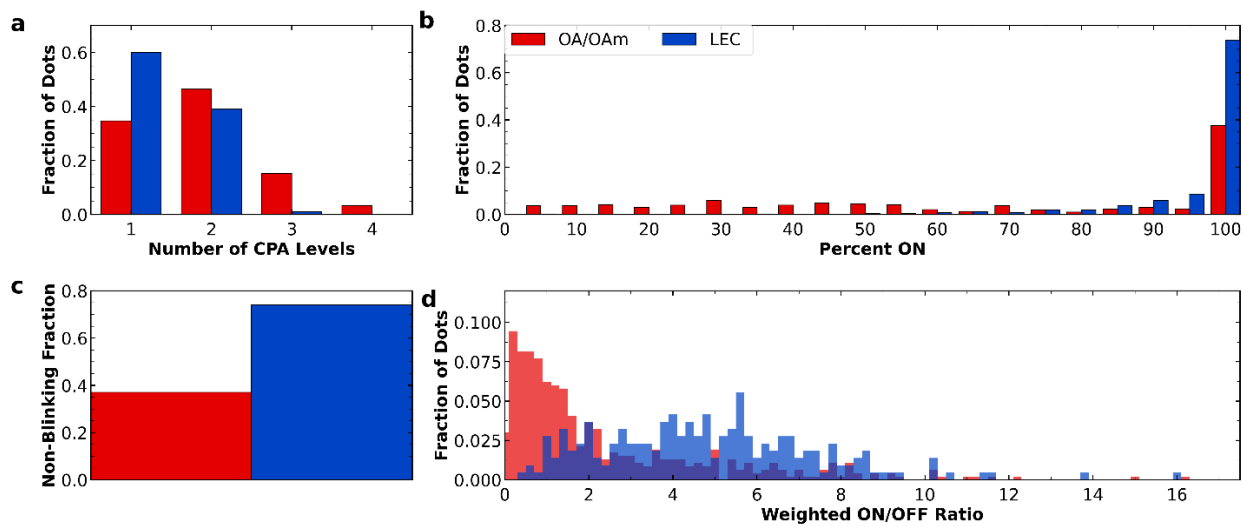

Figure S6. Comparisons of blinking behavior between two additional OA/OAm and lecithin synthetic batches across our four major blinking metrics. These data were obtained at an integration time of 200 ms using a 100x NA 0.95 objective. These statistics are comprised of 716 OA/OAm QDs and 543 lecithin QDs. **a)** number of intensity levels fit to each blinking trace by CPA. The average CPA levels are 1.8 ± 0.8 and 1.4 ± 0.5 for OA/OAm and lecithin respectively **b)** percent of the trace duration that each QD spends ON. The average ON percentages are 44 ± 28% and 87 ± 13% for OA/OAm and lecithin respectively **c)** non-blinking fraction of QDs in each batch, 0.38 and 0.73 for OA/OAm and lecithin respectively and **d)** the weighted ratio of time spent ON to time spent OFF

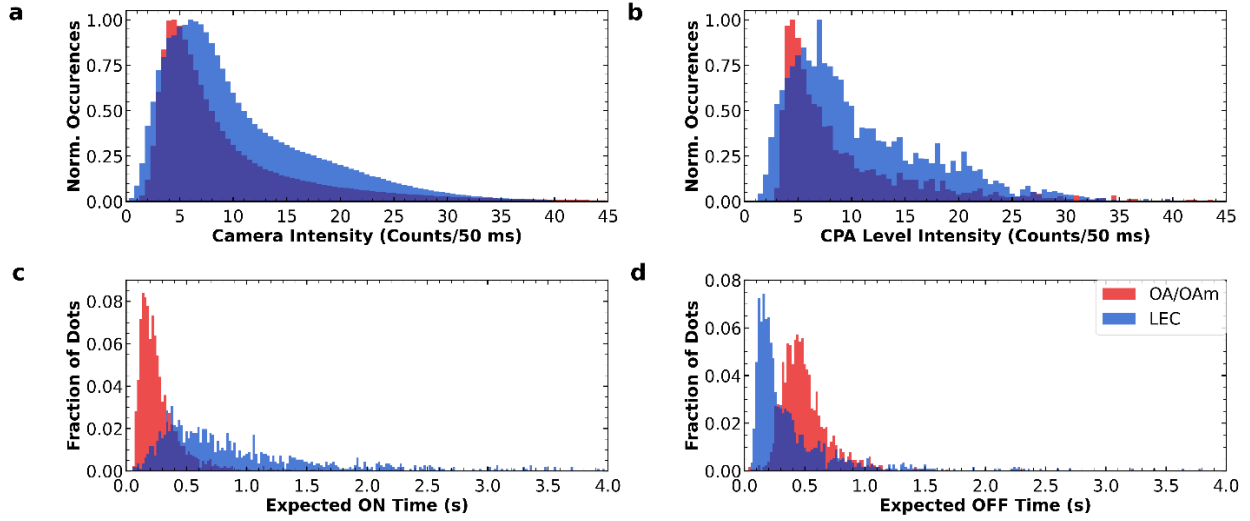

Figure S7. Comparison of blinking behavior between the OA/OAm- and lecithin- capped QDs across four additional metrics **a)** distribution of intensities in the blinking traces according to the camera **b)** distribution of intensities in the blinking traces according to the CPA fits **c)** distribution of the expected ON times from power-law dynamics and **d)** distribution of the expected OFF times from power-law dynamics.

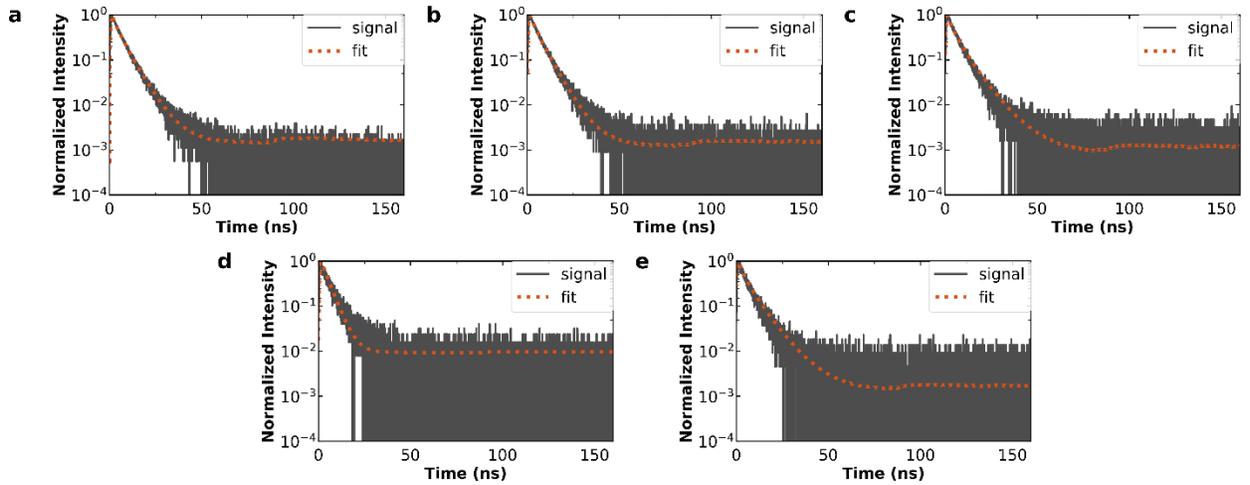

Figure S8. TRPL stretch-exponential fits (eq S2) for LEC 1 QDs as a function of concentration (see Table S1) **a)** solution 1: $\tau_{str}$ = 4.43 ns, $\beta$ = 0.78, $R^2$ = 0.994 **b)** solution 2: $\tau_{str}$ = 4.71 ns, $\beta$ = 0.87, $R^2$ = 0.998 **c)** solution 3: $\tau_{str}$ = 4.52 ns, $\beta$ = 0.74, $R^2$ = 0.985 **d)** solution 4: $\tau_{str}$ = 4.90 ns, $\beta$ = 0.78, $R^2$ = 0.960 and **e)** solution 9: $\tau_{str}$ = 4.50 ns, $\beta$ = 0.76, $R^2$ = 0.982

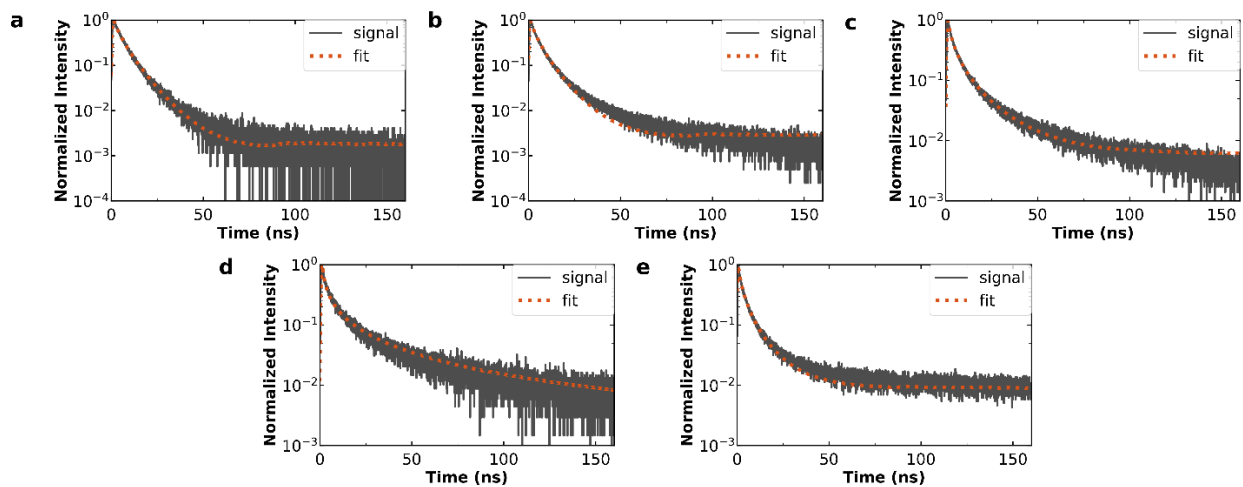

Figure S9. TRPL stretch-exponential fits (eq S2) for OA/OAm QDs as a function of concentration (see Table S1) **a)** solution 1: $\tau_{str}$ = 4.95 ns, $\beta$ = 0.78, $R^2$ = 0.998 **b)** solution 2: $\tau_{str}$ = 4.08 ns, $\beta$ = 0.67, $R^2$ = 0.985 **c)** solution 3: $\tau_{str}$ = 2.81 ns, $\beta$ = 0.45, $R^2$ = 0.993 **d)** solution 4: $\tau_{str}$ = 2.51 ns, $\beta$ = 0.43, $R^2$ = 0.980 and **e)** solution 9: $\tau_{str}$ = 2.08 ns, $\beta$ = 0.45, $R^2$ = 0.992

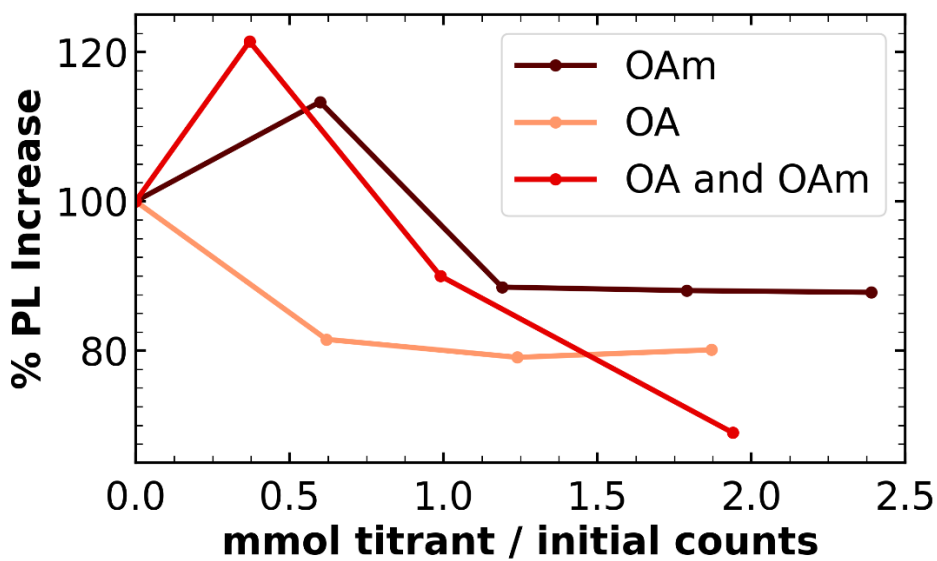

Figure S10. PL increase resulting from the addition of excess ligand to a widefield PL concentration solution of OA/OAm-CsPbBr$_3$ QDs comparing the addition of neat OAm to neat OA and a 1:1 (v:v) mixture of OA and OAm

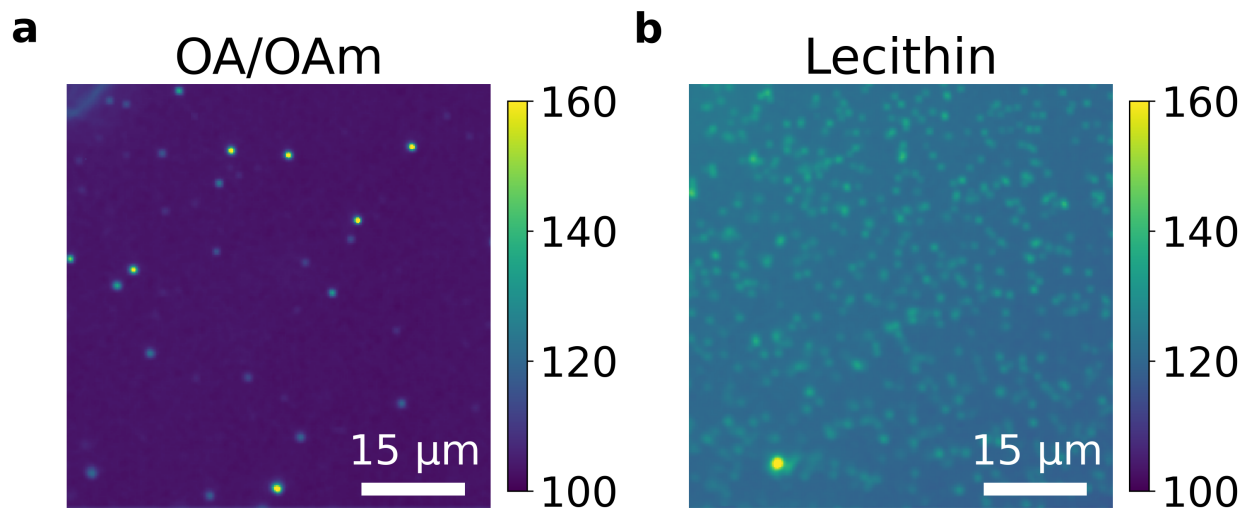

**Figure S11.** Widefield PL images of **a)** OA/OAm QDs and **b)** lecithin QDs. The sample solutions were prepared identically via serial dilution from stock solutions of the same optical density. The images correspond to samples made from solutions B$_{OA/OAm}$ and B$_{lec}$ respectively.

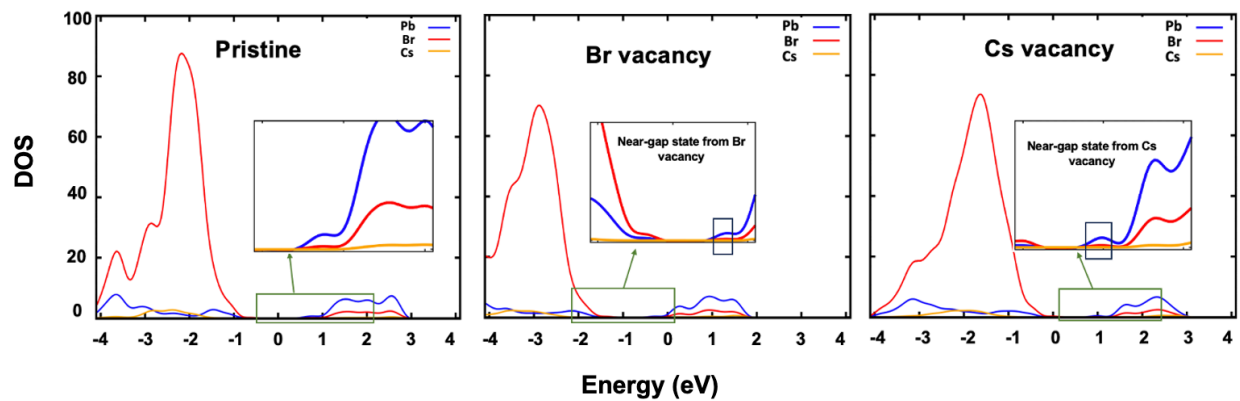

**Figure S12. Density of states (DOS) of Cs-terminated slabs of CsPbBr$_3$.** Calculations were performed at the PBE level with spin orbit coupling (SOC) effects on top of PBE+D3-relaxed structures of the pristine slab (a) as well as slabs containing v•$_{Br}$ (b) and v'$_{Cs}$ (c). Near gap states due to the presence of v•Br and v'Cs are observed and marked with rectangles in the insets of panels b and c.

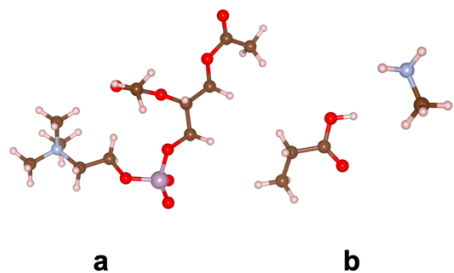

**a**        **b**

**Figure S13. Simplified/truncated forms of the bidentate ligand and monodentate ligand pair used for computational purposes**. Truncated lecithin (a), methylamine and propionic acid (b).

**Table S2.** Calculated binding energies (eV), including solvent effects, corresponding to the single (OAm or OA) and binary monodentate (OAm/OA) ligands passivating the pristine surface. OAm and OA correspond to oleylamine and oleic acid binding the OA/OAm-passivated pristine slab, while OAm* and OA* represent oleylamine and oleic acid binding the OA$^-$/OAm$^+$-passivated defective surface.

| Passivating ligand(s) | OA/OAm | OAm | OA | OAm* | OA* |
|---|---|---|---|---|---|
| Solvated Binding Energy (eV) | -0.42 | -0.17 | -0.65 | -0.25 | -0.67 |

**Table S3.** Calculated binding energies (eV) corresponding to the binary monodentate (OAm$^+$/OA$^-$) and the bidentate ligand (lecithin) passivating different types of surface vacancy pairs (shown in Main Text Figure 5d)

| Vacancy type | **Type *i*** | **Type *ii*** | **Type *iii*** | **Type *iv*** |
|---|---|---|---|---|
| OAm$^+$/OA$^-$ | -1.95 | -1.88 | -1.91 | -2.07 |
| Lecithin | -3.22 | -3.17 | -2.96 | -2.40 |

**Reaction Mechanism and Binding Energy (BE) Calculation**

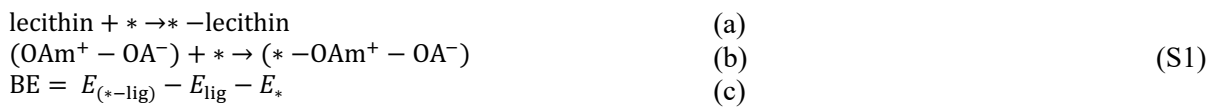

$$\text{lecithin} + * \rightarrow * - \text{lecithin} \quad \text{(a)}$$
$$(\text{OAm}^+ - \text{OA}^-) + * \rightarrow (* - \text{OAm}^+ - \text{OA}^-) \quad \text{(b)} \quad \quad \text{(S1)}$$
$$\text{BE} = E_{(*-\text{lig})} - E_{\text{lig}} - E_* \quad \text{(c)}$$

To compare energetics of lecithin and (OAm$^+$-OA$^-$) binding the QD surface, we considered comparable binding mechanisms (Equations S1a and S1b) whereby comparatively to the zwitterionic ligand, the monodentate ligand pair (OAm$^+$-OA$^-$) binds the quantum dot surface (* in Equation S1) as a complex. The corresponding binding energies can therefore be calculated per Equation S1c, where $E_*$ and $E_{(*-\text{lig})}$ are DFT energies of the bare and ligand-passivated slabs and $E_{\text{lig}}$ represents the energy of either lecithin or the (OAm$^+$-OA$^-$) complex.

*Solvation Effects on Ligand Binding the QD Surfaces*

In principle, the rate of reactions can be controlled by the solvent's polarity, and therefore its dielectric constant, as well as its acceptor (AN) or donor number (DN). To account for the effect of ACN and acetone which are both polar aprotic solvents and used in the washing process of our synthesized QDs, we employed an explicit solvent model.

$$E_{\text{solvn}}{}^{x*\text{solv}} = E_{(*-\text{ligand})}{}^{x*\text{solv}} - E_{(*-\text{ligand})} - x.E_{\text{solv}} \quad (S2)$$
$$\text{BE}|_{\text{solv}} = \text{BE} + (E_{\text{solvn}})_{(*-\text{lig})} - (E_{\text{solvn}})_{\text{lig}} - (E_{\text{solvn}})_* \quad (S3)$$

We used periodic solvated slab models including a layer of liquid solvents on top of the slabs that were used for binding energy calculations. We first attained the number of solvent molecules required to sufficiently solvate the ligands as well as bare and passivated slabs, by gradually adding solvent molecules ($x$) to the system and calculating the solvation energy ($E_{\text{solvn}}{}^{x*\text{solv}}$) per Equation S2. In this equation, $x.E_{\text{solv}}$, represents the energy of $x$ solvent molecules in liquid phase, and $E_{(*-\text{ligand})}$ and $E_{(*-\text{ligand})}{}^{x*\text{solv}}$ are total energies of the passivated slab without solvent and containing $x$ solvent molecules, respectively. As a representative, the onset of the plateau in Figures S13b and S13c shows the minimum number of solvent molecules required to sufficiently solvate the (OAm$^+$-OA$^-$)-passivated and lecithin-passivated slabs of CsPbBr$_3$. Accordingly, we used 10 and 20 solvents for the adjacent and non-adjacent (twice as large unit cell) vacancy cases, respectively. Finally, to obtain the net effect of solvation on the binding energy, we calculated the solvation energy for the ligand ($E_{\text{solvn}}$)$_{\text{lig}}$, clean slab ($E_{\text{solvn}}$)$_*$ and the ligand-passivated slab ($E_{\text{solvn}}$)$_{(*-\text{lig})}$, and computed the binding energy in the presence of solvents (BE |$_{\text{solv}}$) by applying the obtained net solvation energy to the BE, per Equation S3.

**Binding Energy Calculation of Mono-ligand Passivated Single-vacancy Surfaces**

In addition to vacancy pairs, we studied Br$^-$-rich and OAm$^+$-rich conditions where the single vacancy surface is passivated by a monodentate ligand. We calculated the BE of OAm$^+$ passivating CsPbBr$_3$ by gradually pulling the ligand off the surface and obtained the corresponding energy *vs*. distance plot (Figure S13a). $E_d$ is the energy required to pull the ligand off the surface by a distance of $d$ (Å), marked on the x-axis, $d$ = 0 (Å) refers to the ligand binding the surface, and $d$ = -1 (Å) corresponds to the ligand filling the vacant site. It can be seen that OAm$^+$ strongly binds the surface (BE = -4.1 eV) which can primarily be attributed to the size of the ligand and therefore, the capability of the ammonium moiety to access and occupy the vacant site (v'$_{\text{Cs}}$).

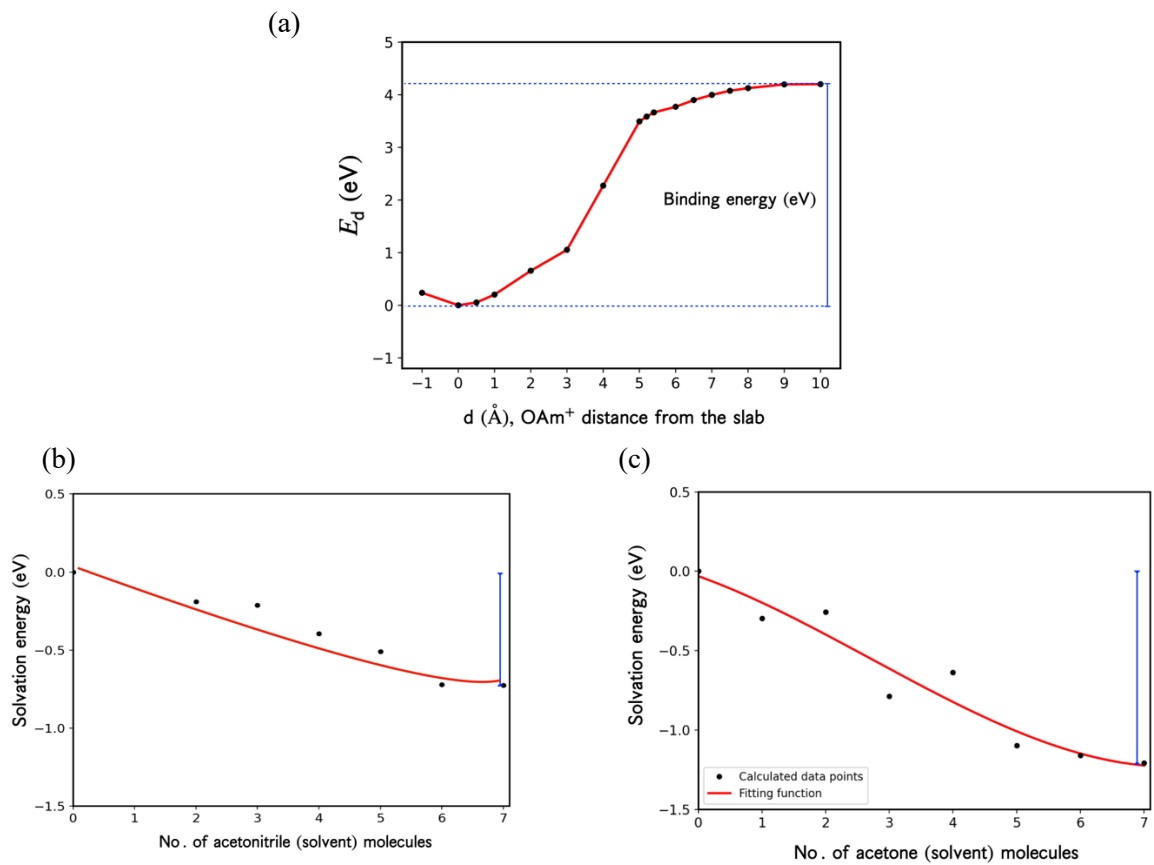

**Figure S14.** (a) Binding energy calculation of the OAm$^+$-passivated slab containing a single Cs vacancy (v'$_{Cs}$). $E_d$ is the energy required to pull the ligand off the surface by $d$ (Å) (shown on the x-axis), $d = 0$ (Å) refers to the ligand binding the surface, and $d = -1$ (Å) corresponds to the ligand filling the vacant site. (b) Solvation energies versus the number of solvent molecules, calculated per Equation S3, for (OAm$^+$-OA$^-$)-passivated and (c) lecithin-passivated slabs of CsPbBr$_3$ are shown. Solid blue lines indicate the solvation energies at the onset of the plateau corresponding to the number of solvent molecules sufficient to solvate the passivated slabs.

**Equilibrium Constants and Binding Energies**

To better understand the calculated binding energies in the context of the system – we compute the ligand binding equilibrium constant from both the DFT binding energies (Main Text Figure 5d) and the experimental PL dilution results (Main Text Figure 4c).

The ligand binding equilibrium constant from DFT was calculated with equation S4:

$$\Delta G° = -RT\ln(K_{ligand}) \qquad (S4)$$

where $\Delta G°$ is the binding energy in J/mol, R is the gas constant (8.314 J mol$^{-1}$ K$^{-1}$), T is the temperature and $K_{ligand}$ is the ligand binding equilibrium constant.

The experimental ligand binding equilibrium constant was calculated through a series of steps. First to determine the ligand and nanocrystal solutions we dried 2 mL of a 290 nM solution of the OA/OAm-capped QDs in a centrifuge tube of known weight and measured the total QD and ligand mass to be 3 ± 1 mg. The total mass contribution from the OA/OAm-capped QDs was calculated to be 1 mg so the maximum ligand contribution was 2 mg. The dry mass of 2 mL of a 140 nM solution of lecithin-capped QDs was measured to be 2 ± 1 mg. The total mass contribution from the lecithin-capped QDs was calculated to be 0.3 mg so the maximum ligand contribution was 1.7 mg.

Assuming 100 binding sites per QD and that the QDs start as completely bound, the initial concentrations at the drop-off point (solution 6, 0.04 nM) were calculated to be [OA/OAm-QD] = 3.6E-11 M, [free OA/OAm] = 1.9E-7 M, [lecithin-QD] = 3.6E-11 M and [free lecithin] = 1.3E-7 M.

Based on Figure 4c the OA/OAm the integrated PL intensity at 0.04 nM is ~7 times lower than expected, meaning that at equilibrium only 15% of the QDs are still bound by ligand and emissive. For lecithin the PL intensity is nearly linear and an estimated 99% of the QDs are still bound by ligand and emissive. From these numbers $K_{ligand}$ can be calculated using equation S5

$$K_{ligand} = \frac{f_{emissive}[\text{QD-ligand}]}{(1-f_{emssive})[\text{QD-ligand}][\text{free ligand}]} \tag{S5}$$

Where $f_{emissive}$ is the fraction of quantum dots which are still bound and emissive, [QD-ligand] is the starting concentration of fully bound QDs in solution 6 and [free-ligand] is the starting concentration of free ligand in solution 6.

**Table S4.** Ligand binding equilibrium constant $K_{ligand}$ for solvated OA/OAm and lecithin QDs computed from DFT and experimental results

|  |  | $K_{ligand}$ |
|---|---|---|
| **DFT** | *OA/OAm* | 3.7 E4 |
|  | *lecithin* | 1.3 E34 |
| **Experimental** | *OA/OAm* | 9.1 ± 0.3 E5 |
|  | *lecithin* | 7.4 ± 0.6 E8 |

Using equation S4 the experimental binding energy of OA/OAm to $CsPbBr_3$ is -0.34 eV. For lecithin the experimental binding energy is -0.50 eV. The observed discrepancy between our theoretical and experimental binding energies for lecithin can likely be attributed to our decision to use a truncated version of the molecule for computational feasibility. We anticipate that the solvent effects would have been more pronounced, aligning the calculated binding energies more closely with experimental values, had the full molecular structure been incorporated in our analysis.

Table S5. Serial dilutions resulting in QD solutions used for concentration dependent PL. Solutions used for TRPL are indicated with a star.

|  | Volume of OA/OAm QD Solution | Concentration of OA/OAm QDs (mg/mL) | Volume of lecithin QDs | Concentration of lecithin QDs (mg/mL) | Volume of Hexanes (mL) |
|---|---|---|---|---|---|
| **Solution 1 (Ensemble PLQY Concentration)*** | 100 µL of stock solution | 3.3E1 | 200 µL of stock solution | 6.3E-1 | 4 |
| **Solution 2*** | 20 µL of stock solution | 6.6E0 | 40 µL of stock solution | 1.3E-1 | 4 |
| **Solution 3*** | 10 µL of stock solution | 3.3E0 | 20 µL of stock solution | 6.3E-2 | 4 |
| **Solution 4*** | 2 µL of stock solution | 6.6E-1 | 4 µL of stock solution | 1.3E-2 | 4 |
| **Solution 5** | 1 µL of stock solution | 3.3E-1 | 2 µL of stock solution | 6.3E-3 | 4 |
| **Solution 6** | 200 µL of solution 3 | 1.7E-1 | 200 µL of solution 3 | 3.2E-3 | 4 |
| **Solution 7** | 133 µL of solution 3 | 1.1E-1 | 133 µL of solution 3 | 2.1E-3 | 4 |
| **Solution 8** | 100 µL of solution 3 | 8.3E-2 | 100 µL of solution 3 | 1.6E-3 | 4 |
| **Solution 9 (Microscopy Concentration)*** | 400 µL of solution 4 | 6.6E-2 | 400 µL of solution 4 | 1.3E-3 | 4 |
| **Solution 10** | 333 µL of solution 4 | 5.5E-2 | 333 µL of solution 4 | 1.1E-3 | 4 |

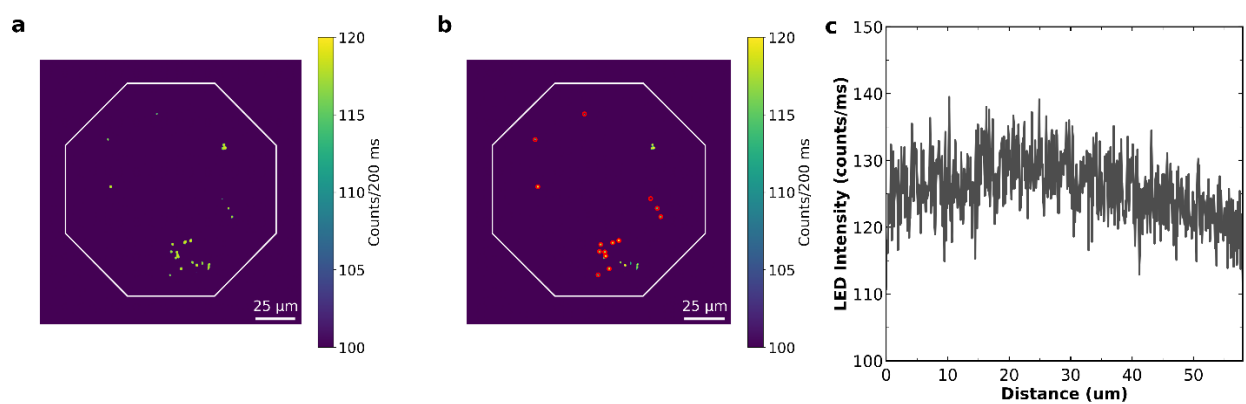

**Figure S15. a)** quantum dot sample illuminated with the field-stop (white octagon) partially closed to demonstrate the quality of particle selection. **b)** quantum dot sample illuminated with the field-stop (white octagon) partially closed and selected particles circled in red to demonstrate the quality of particle selection. The particle selection algorithm accounts for particle brightness, size and nearest neighbor distance. **c)** Illumination homogeneity of the LED across the objective flat field of view. Y-axis limits represent ±20% of the mean illumination intensity.

| | CPA Identified Intensity States | | | | | | |
|---|---|---|---|---|---|---|---|
| | 1 | 2 | 3 | 4 | 5 | 6 | 7 |
| 1 | 1.00 | 0.00 | 0.00 | 0.00 | 0.00 | 0.00 | 0.00 |
| 2 | 0.00 | 0.84 | 0.16 | 0.00 | 0.00 | 0.00 | 0.00 |
| 3 | 0.00 | 0.00 | 1.00 | 0.00 | 0.00 | 0.00 | 0.00 |
| 4 | 0.00 | 0.00 | 0.00 | 1.00 | 0.00 | 0.00 | 0.00 |
| 5 | 0.00 | 0.00 | 0.00 | 0.08 | 0.92 | 0.00 | 0.00 |
| 6 | 0.00 | 0.00 | 0.04 | 0.18 | 0.44 | 0.35 | 0.00 |
| 7 | 0.00 | 0.00 | 0.96 | 0.04 | 0.00 | 0.00 | 0.00 |

(True Number of Intensity States on vertical axis)

Figure S16. Confusion matrix quantifying the performance of our custom CPA package. Confusion matrix was generated by creating synthetic blinking traces with a power law exponent of α = 1.5, a time step size of 50 ms and $n$ states evenly spaced between 100 and 140 counts/50 ms. 100 synthetic traces[1] were generated for each $n$ in the range of 1 to 7 (the true number of states). The synthetic traces were analyzed using our CPA package and the number of fit states was extracted (the number of CPA determined states). This test shows that we can reliably resolve up to 5 intensity states in blinking traces at the experimental conditions (time step size = 50 ms and intensity range of 100 to 140 counts/50 ms).

**Change Point Analysis to Blinking Statistics**

Our analysis package uses CPA to determine the number of intensity levels in each trace and which intensity level each time point belongs to.[2,3] From this information we calculate the dwell time probability distribution for each level and fit to a power law[4] (Equation S6) and a truncated power law[5] (Equation S7). The fit that yields a larger $R^2$ value is used for further analysis of the intensity level.

$$P(t) = Ct^{-\alpha} \tag{S6}$$

$$P(t) = Ct^{-\alpha} e^{t/T_c} \tag{S7}$$

Where $P(t)$ is the dwell time probability distribution, $C$ is an exponential pre-factor, $t$ is the dwell time, $\alpha$ is the power law exponent and $T_c$ is the cutoff time. From these fits the dwell time expectation value $\langle t \rangle$ can be calculated according to Equations S8 (power law) or S9 (truncated power law).

$$\langle t \rangle = \left(\frac{\alpha+1}{\alpha+2}\right)\left(\frac{t_{max}^{\alpha+2} - t_{min}^{\alpha+2}}{t_{max}^{\alpha+1} - t_{min}^{\alpha+1}}\right) \tag{S8}$$

$$\langle t \rangle = -T_c \left(\frac{\Gamma\left(\alpha+2, \frac{-t_{max}}{T_c}\right) - \Gamma\left(\alpha+2, \frac{-t_{min}}{T_c}\right)}{\Gamma\left(\alpha+1, \frac{-t_{max}}{T_c}\right) - \Gamma\left(\alpha+1, \frac{-t_{min}}{T_c}\right)}\right) \tag{S9}$$

Where $t_{max}$ is the maximum observed dwell time, $t_{min}$ is the minimum observed dwell time and $\Gamma$ is the upper incomplete gamma function.

For the purposes of the post-CPA analysis an ON-state is defined as the most intense CPA identified level and all other CPA identified intensity levels are defined as OFF-states. From the above information we can calculate our blinking statistics.

$$ON\ \% = \frac{T_{ON}}{T_{total}} * 100 \quad \text{(S10)}$$

$$\tau_{ON} = \langle t \rangle_{ON} \quad \text{(S11)}$$

$$\tau_{OFF} = \sum_n \frac{T_n}{(T_{total} - T_{ON})} \langle t \rangle_n \quad \text{(S12)}$$

$$\tau_{ON/OFF} = \frac{T_{ON} \tau_{ON}}{(T_{total} - T_{ON}) \tau_{OFF}} \quad \text{(S13)}$$

Where $T_{ON}$ is the total dwell time of the ON-state, $T_{total}$ is the total measurement time, $n$ is the number of OFF states, $T_n$ are the total dwell times of each OFF-state, $\tau_{ON}$ is the expected ON event dwell time (Figures S5c and S7c), $\tau_{OFF}$ is the weighted expected OFF event dwell time (Figure Figures S5d and S7d) and $\tau_{ON/OFF}$ is the weighted ON/OFF ratio (Figure 3d).